\documentclass[
 reprint,
superscriptaddress,
amsmath,amssymb,
pra
]{revtex4-2}

\usepackage{graphicx}
\usepackage{dcolumn}
\usepackage{bm}
\usepackage[mathlines]{lineno}

\usepackage{mathtools} 
\usepackage{amsthm}

\usepackage{multirow}
\usepackage{array}
\usepackage{makecell}

\usepackage[colorlinks]{hyperref}
\hypersetup{linkcolor=magenta,urlcolor=blue,citecolor=blue,pdfstartview={FitH},hyperfootnotes=false,unicode=true}

\def\be{\begin{equation}}
\def\ee{\end{equation}}
\def\bea{\begin{eqnarray}}
\def\eea{\end{eqnarray}}
\def\nn{\nonumber}

\def\d{{\rm d}}

\newcommand{\ket}[1]{\left|#1\right\rangle} 
\newcommand{\bra}[1]{\left\langle #1\right|} 
\newcommand{\braket}[1]{\left\langle #1 \right\rangle} 
\newcommand{\Tr}[1]{\operatorname{\textnormal{Tr}}\left( {#1} \right)} 
\newcommand{\Trs}[1]{\operatorname{\textnormal{Tr}}^2\left( {#1} \right)}

\newcommand{\set}[1]{\left\{ #1 \right\}} 
\newcommand{\abs}[1]{{\left| #1 \right|}}

\newcommand{\norm}[1]{\left\lVert#1\right\rVert}

\newcommand{\sbraket}[1]{\left[ #1 \right]}

\begin{document}

\title{Coherent Quantum Speed Limits}

\author{Xuhui Xiao}
\email{These authors contribute equally to this work}
\affiliation{School of Physics Science and Engineering, Tongji University, Shanghai 200092, China}

\author{Hai Wang}
\email{These authors contribute equally to this work}
\affiliation{School of Mathematics and Statistics, Nanjing University of Science and Technology, Nanjing 210094, Jiangsu, China}

\author{Xingze Qiu}
\email{xingze@tongji.edu.cn}
\affiliation{School of Physics Science and Engineering, Tongji University, Shanghai 200092, China}

\date{\today}

\begin{abstract}
We establish a comprehensive theoretical framework for coherent quantum speed limits (QSLs), deriving fundamental bounds on the rate of quantum evolution that explicitly isolate the contribution of quantum coherence. By applying H\"older's inequality for matrix norms to the Liouville-von Neumann equation, we construct two infinite families of QSLs for general unitary dynamics. These bounds are characterized by coherence measures based on Schatten $p$-norms and Hellinger distance, respectively, defined with respect to the instantaneous energy eigenbasis. Unlike traditional Mandelstam-Tamm bounds, our approach disentangles the quantum state's coherence structure from the Hamiltonian's energy scale. Using the Landau-Zener model accelerated by shortcuts to adiabaticity, we demonstrate that coherence functions as a critical kinematic resource: achieving faster evolution entails maintaining a state with high coherence relative to the instantaneous basis. Our results provide a resource-theoretic perspective on time-energy uncertainty, offering insights into the fundamental limits of quantum control and information processing.

\end{abstract}

\maketitle

\section{Introduction}

The fundamental limits imposed by quantum mechanics on the time evolution of physical systems constitute a cornerstone of modern theoretical physics \cite{Frey2016, Deffner_2017}. Central to this inquiry is the concept of the quantum speed limit (QSL), which defines the minimum time $T_{\rm QSL}$ required for a quantum system to evolve between two distinguishable states. Originally conceived as a rigorous manifestation of the time-energy uncertainty principle, QSLs have evolved from foundational curiosities into practical tools essential for quantum communication \cite{Bekenstein1981PRL, Deffner2020PRR}, quantum computation \cite{Lloyd2000, Deffner_2021}, quantum metrology \cite{Giovannetti2011NatPhoton, delCampo2017PRL}, quantum optimal control \cite{Caneva2009PRL, Brody2015PRL, Campbell2017PRL}, quantum information \cite{Campaioli_2022_NJP, Mohan_2022_NJP}, as well as nonequilibrium quantum thermodynamics \cite{Deffner2010PRL, Campaioli2017PRL, Hasegawa2023NC, Gyhm_2024_PRA} and many-body physics \cite{Fogarty2020PRL, Girolami2021PRL}.

The genesis of QSL theory traces back to the seminal work of Mandelstam and Tamm (MT) in 1945 \cite{MTBound_1945}. They derived a bound for unitary dynamics generated by a time-independent Hamiltonian $H$, relating the minimum evolution time to the energy uncertainty $\Delta H = \sqrt{\langle H^2\rangle-\langle H\rangle^2}$: $T \geq T_{\rm MT}:= \pi /(2 \Delta H)$ (units are such that $\hbar=1$). 
Half a century later, Margolus and Levitin (ML) provided a complementary bound based on the mean energy relative to the ground state, $\bar{H} = \braket{H} - E_g$: $T \geq T_{\rm ML}:= \pi/(2 \bar{H})$ \cite{Margolus1998}. 
For general time-independent systems, the unified bound is the maximum of the two: $T_{\rm QSL} = \max\{T_{\rm MT}, T_{\rm ML}\}$ \cite{Levitin2009PRL}. These results have been extensively generalized to time-dependent Hamiltonians \cite{Anandan1990PRL, Pfeifer1993PRL}, mixed states \cite{Uhlmann1992, Marvian2016PRA, Pires2016PRX, Campaioli2018PRL, Niklas_2022_NJP, Alberto_2024_PLA}, open quantum systems \cite{Deffner2013PRL, delCampo2013PRL, Taddei2013PRL, Marvian2015PRL}, multi-partite entangled systems \cite{Vittorio2003PRA, Zander_2007}, and the evolution of observables \cite{delCampo2022PRX, Hamazaki2022PRXQuantum}. 
Furthermore, speed limits have been found to exist even for classical dynamics \cite{Shanahan2018PRL, Okuyama2018PRL, Shiraishi2018PRL, Nicholson2020NP}. 
Experimentally, QSLs have been demonstrated in numerous platforms, such as optical cavities \cite{Cimmarusti2015PRL} and cold atoms \cite{Lam2021PRX, Gal2021SciAdv}.

In recent years, quantum coherence has emerged as a central pillar of quantum resource theories \cite{Plenio2017RMP}, alongside entanglement \cite{Chitambar_2016_PRL} and correlations \cite{Tan_2016_PRA}.
Coherence quantifies the superposition of a quantum state with respect to a preferred basis---in the context of dynamics, this is naturally the instantaneous energy eigenbasis. 
If a time-independent system is perfectly incoherent (diagonal) in the energy basis, it is stationary; it cannot evolve. 
Conversely, fast dynamics require significant superposition among energy eigenstates. 
While the connection between coherence and speed is intuitive and has been explored in various contexts \cite{Pires2015PRA, Marvian2016PRA, Pires2016PRX, delCampo_2016_SRep, delCampo_2017_PRL, Rossatto2020PRA, Mohan_2022_NJP, Paulson_2022}, existing generalizations often incorporate coherence implicitly. 
For instance, bounds based on the Wigner-Yanase Skew Information (WYSI) \cite{Pires2015PRA, Marvian2016PRA, Pires2016PRX} combine geometric properties of the state with the Hamiltonian variance in a single metric. 
This conflation makes it difficult to distinguish whether a speedup arises from a change in the state's structure (its coherence) or simply from an increase in the energy scale of the Hamiltonian. 
Explicitly separating the ``amount of coherence" from the Hamiltonian's spectral properties is necessary for a transparent resource-theoretic understanding of fast quantum evolution.

\begin{table*}
    \centering
    \caption{Comparisons between our coherent QSLs and the corresponding established results. The latter are summarized as follows: 
(i) $T_{\rm AA}$: Anandan and Aharonov generalized the original MT-bound to time-dependent system by using the fact that Bures angle is the geodesic length in the state space \cite{Anandan1990PRL}. 
When restricting to the time-independent case, $T_{\rm AA}$ reduces to $\tilde{T}_{\rm AA}$. 
(ii) $T_{\Theta}$: In Ref.~\onlinecite{Campaioli2018PRL}, the authors utilized the generalized Bloch angle $\Theta(\rho,\sigma)=\arccos[(\Tr{\rho\sigma}-1/N)/(\Tr{\rho^2}-1/N)]$ to derive the bound $T_{\Theta}$. 
(iii) $T_{\rm WY}$: In Ref.~\onlinecite{Pires2016PRX}, the authors studied the geometric bound $T_{\rm WY}$ based on the WYSI metric. 
Here, $N$ is the dimension of the Hilbert space, $\mathcal{L} = \arccos\sqrt{F}$ is the Bures angle with $F$ the quantum fidelity, $\Delta H$ is the energy uncertainty, $\mathcal{L}_{\rm WY}=\arccos[F_{\rm A}]$ is the WYSI metric with $F_{\rm A}$ the quantum affinity, $I$ is the WYSI, and $C_p$ ($p=1,2$) and $C_{\rm H}$ are the coherence measures based on Schatten $p$-norm and Hellinger distance, respectively. 
More details about the notations can be found in the main text. 
    }
    \vspace{5pt}
    \renewcommand\arraystretch{2.5}
    \resizebox{0.95\linewidth}{!}{
        \begin{tabular}{  l | c | c }
        \hline
        \hline
        
        \makecell[l]{ Coherence \\ measure \\[2pt]}    & This work &  Established results  \\
        \hline

        \multirow{3}{*}{\makecell[l]{ \\[50pt] Schatten \\ $p$-norm}}   
       
        & {\makecell[c]{ \\
        {$\displaystyle \tilde{\mathcal{T}}_{\rm S,\, Pure}(1,\infty) = \frac{1-F(\ket{\psi_0},\ket{\psi_T})}{C_1(\ket{\psi_0})\cdot \Delta H(\ket{\psi_0}) }  $} [Eq.~\eqref{eq:Bound_S_MT_p=1}]  \\ [10pt]  
        {$\displaystyle \tilde{\mathcal{T}}_{\rm S,\, Pure}(2,2) = \frac{1-F(\ket{\psi_0},\ket{\psi_T})}{\sqrt{2}\, C_2(\ket{\psi_0}) \cdot \Delta H(\ket{\psi_0})}  $} [Eq.~\eqref{eq:Bound_S_MT_p=2}]
        \\[10pt]  
        }} 
        & {$\displaystyle \tilde{T}_{\rm AA} = \frac{\mathcal{L}(\ket{\psi_0},\ket{\psi_T})}{\Delta H(\ket{\psi_0})}$} \cite{Anandan1990PRL} \\[10pt]  
        \cline{2-3}
        
        & {\makecell[c]{ \\
        {$\displaystyle \mathcal{T}_{\rm S,\, Pure}(1,\infty) = \frac{1-F(\ket{\psi_0},\ket{\psi_T})}{\frac{1}{T}\int^T_0\d t \, C_1(\ket{\psi_t}) \cdot \Delta H_t(\ket{\psi_0})}  $} [Eq.~\eqref{eq:Bound_S_Pure_p=1}]  \\ [10pt]  
        {$\displaystyle \mathcal{T}_{\rm S,\, Pure}(2, 2) = \frac{1-F(\ket{\psi_0},\ket{\psi_T})}{\frac{\sqrt{2}}{T}\int^T_0\d t \, C_2(\ket{\psi_t}) \cdot \Delta H_t(\ket{\psi_0})}  $} [Eq.~\eqref{eq:Bound_S_Pure_p=2}] 
        \\[10pt]  
        }} 
        & {$\displaystyle T_{\rm AA} = \frac{\mathcal{L}(\ket{\psi_0},\ket{\psi_T})}{\frac{1}{T}\int^T_0\d t \, \Delta H_t(\ket{\psi_t})}$} \cite{Anandan1990PRL}  \\[10pt] 
        \cline{2-3}

         & {\makecell[c]{ \\
         {$\displaystyle  \mathcal{T}_{\rm S}(2, 2) 
= \frac{\sbraket{1 - F_{\rm RP}(\rho_0,\rho_T)}\Tr{\rho^2_0}}{\frac{\sqrt{2}}{T}\int^T_0\d t \, C_2(\rho_t) \sqrt{\Tr{\rho^2_0 H^2_t-(\rho_0 H_t)^2}}}  $} [Eq.~\eqref{eq:Bound_S_p=2}] \\[15pt]  
           }}

         & 
         {$\displaystyle T_{\Theta} = \frac{\Theta(\rho_0,\rho_T)}{\frac{\sqrt{2}}{T}\int^T_0\d t \, \sqrt{\frac{\Tr{\rho^2_t H^2_t-(\rho_t H_t)^2}}{\Tr{\rho^2_t}-1/N}} } $} \cite{Campaioli2018PRL} \\[15pt]

        \hline
        \makecell[l]{ Hellinger \\ distance}

        & {\makecell[c]{ \\
         {$\displaystyle \mathcal{T}_{\rm H}(2,2) = \frac{1-F_{\rm A}(\rho_0,\rho_T)}{\frac{\sqrt{2}}{T}\int^T_0\d t \,  \sqrt{C_{\rm H}(\rho_t)}\cdot \sqrt{I(\rho_0,H_t)}}  $} [Eq.~\eqref{eq:Bound_H}] \\[10pt] 
         }}
         
        & {$\displaystyle T_{\rm WY} = \frac{\mathcal{L}_{\rm WY}(\rho_0,\rho_T)}{\frac{\sqrt{2}}{T}\int^T_0\d t \, \sqrt{I(\rho_t,H_t)} } $} \cite{Pires2016PRX}
          \\[10pt]

        \hline 
        \hline
        \end{tabular}
        }
    \label{tab:compare_QSL}
\end{table*}

In this work, we present a unified framework for coherent QSLs by employing H\"older's inequality for matrix norms.
We construct two infinite families of bounds for general unitary dynamics, measuring coherence via Schatten $p$-norms \cite{Plenio2014PRL, deVicente_2017} and the Hellinger distance \cite{Jin2018PRA}, respectively. 
Our approach is versatile enough to provide a unified framework for general unitary dynamics, encompassing both time-dependent and mixed state scenarios. 
As shown in Table~\ref{tab:compare_QSL}, our QSLs clearly disentangle the contribution of the coherence of the evolved state from the energy uncertainty (or WYSI) of the generator, thus clarifying their individual roles.
To validate the theoretical utility of these bounds, we perform exhaustive analyses of the experimentally relevant Landau-Zener (LZ) model \cite{Ivakhnenko_2023} and a related two-qubit model.
We show that our QSLs can provide tighter bounds than established results and are asymptotically saturable in the adiabatic limit.
Moreover, by utilizing shortcuts to adiabaticity (STA) \cite{Berry_2009, Muga_2019_RMP}, we demonstrate that our bounds can also be saturated at finite time.
This analysis explicitly reveals that fast dynamics necessitates a greater amount of coherent superposition of energy eigenstates, thereby identifying coherence as a key kinematic resource for unitary evolution. 
Our findings have immediate implications for exploiting coherence to control dynamics in quantum simulation platforms.


\section{Unified Framework for Coherent Quantum Speed Limits} 
\label{sec:unified}

We present a unified theoretical framework for deriving coherent QSLs. 
We consider general unitary dynamics and determine the minimal time required to evolve the state $\rho_t$ from an initial state $\rho_0$ to a distinct final state under the time-dependent Hamiltonian $H_t$.
Our goal is to disentangle the contribution of coherence from the energetic driving of the Hamiltonian. 
To this end, we introduce a unified framework based on H\"older's inequality for matrix norms, applied to two distinguishable measures: the relative purity $F_{\rm RP}(\rho,\sigma) = \Tr{\rho\sigma}/\Tr{\rho^2}$ \cite{Audenaert2014} and the Hellinger distance $D_{\rm H}(\rho,\sigma) \coloneqq {\rm Tr}[(\sqrt{\rho}-\sqrt{\sigma})^2] = 2- 2F_{\rm A}(\rho,\sigma)$, with $F_{\rm A}(\rho,\sigma)={\rm Tr}(\sqrt{\rho}\sqrt{\sigma})$ denoting the quantum affinity \cite{Luo2004PRA}. 
We prioritize relative purity for its analytical tractability when deriving bounds using norm inequalities (like Hölder's inequality) directly on the Liouville equation, which allows for a transparent factorization of the quantum coherence from the QSLs. Additionally, for pure states, it reduces exactly to the standard quantum fidelity, ensuring natural generalization. 
Similarly, employing the Hellinger distance allows us to establish a direct link to the corresponding coherence measure and connects our results with the information-theoretic quantity WYSI. 
Previous QSL studies have successfully employed relative purity and quantum affinity to characterize distinguishability in open and closed system dynamics \cite{delCampo2013PRL} and to investigate contexts involving information geometry \cite{Pires2016PRX}, respectively.

The central ingredient in our framework is the instantaneous incoherent reference state. At any time $t$, the Hamiltonian $H_t$ defines a preferred basis, the instantaneous energy eigenbasis $\set{\ket{n_t}}$ such that $H_t\ket{n_t}=E_{n,t}\ket{n_t}$.
We define the set of incoherent states $\mathcal{I}_t$ as the set of all density matrices $\sigma_t$ that are diagonal in this instantaneous basis, i.e., $[\sigma_t, H_t]=0$.
The coherence of the evolved state $\rho_t$ is then measured by its minimum distance to this set, defined as $C(\rho_t)=\min_{\sigma_t \in \mathcal{I}_t}  D(\rho_t,\sigma_t)$, where $D$ is a valid distance measure, in accordance with the seminal framework established in Ref.~\onlinecite{Plenio2014PRL}. 
This rigorous definition ensures that our framework relies on standard concepts in quantum resource theory.
Consequently, our coherence measures are time-dependent quantities measuring the distance of the evolved state $\rho_t$ from the instantaneously changing set $\mathcal{I}_t$.

\subsection{Quantum speed limits based on Schatten norms}
We first consider the relative purity. The dynamics of $\rho_t$ is governed by the Liouville-von Neumann equation $\d\rho_t/\d t=i[\rho_t,H_t]$.
The rate of change of relative purity is given by $-\d F_{\rm RP}(\rho_0,\rho_t)/\d t = -i\Tr{\rho_0[\rho_t,H_t]}/\Tr{\rho^2_0}$.
By introducing the incoherent reference state $\sigma_t \in \mathcal{I}_t$ and applying H\"older's inequality, we obtain the fundamental inequality (see Appendix \ref{appsec:Derivation} for detailed derivation):
\be
-\frac{\d F_{\rm RP}(\rho_0,\rho_t)}{\d t} \leq \frac{C_p(\rho_t)  \norm{ [H_t,\rho_0] }_q}{\Tr{\rho^2_0}} \,.
\label{eq:diff_inequality_S}
\ee
Here, 
\be
C_p(\rho_t)= \min_{\sigma_t \in \mathcal{I}_t}  \norm{ \rho_t -  \sigma_t }_p
\label{eq:C_p}
\ee
is the coherence measure \cite{Plenio2014PRL, deVicente_2017} based on the Schatten $p$-norm $\norm{A}_p= \sbraket{\Tr{|A|^p}}^{1/p}$ ($p\in[1,\infty]$) with $1/p+1/q=1$. 
Integrating both sides yields
\be
\frac{1-F_{\rm RP}(\rho_0,\rho_T)}{T}
\leq \frac{1}{T}\int^T_0\d t \frac{C_p(\rho_t)  \norm{ [H_t,\rho_0] }_q}{\Tr{\rho^2_0}}\, ,
\label{eq:F_int}
\ee
where we have used the fact that $F_{\rm RP}(\rho_0,\rho_0)=1$. 
Consequently, the time $T$ required to evolve from an initial state $\rho_0$ to some final state $\rho_T$ is bounded by
\be
T \geq \mathcal{T}_{\rm S}(p,q) 
\coloneqq \frac{[1-F_{\rm RP}(\rho_0,\rho_T)]\Tr{\rho^2_0}}{\frac{1}{T}\int^T_0\d t \, C_p(\rho_t) \norm{ [H_t,\rho_0] }_q} \, ,
\label{eq:Bound_S_pq}
\ee
which holds for all $p,\,q \in[1,\infty]$ satisfying $1/p+1/q=1$.

The infinite family of QSLs $\mathcal{T}_{\rm S}(p,q)$ constitutes our first main result. 
It applies to any unitary dynamics and establishes a one-to-one correspondence with the coherence measure based on the Schatten $p$-norm. 
These bounds explicitly highlight the contribution of the coherence of the evolved state $C_p(\rho_t)$, thereby cementing the role of coherence as a key resource in driving quantum dynamics and establishing a rigorous trade-off between speed, coherence, and energy uncertainty. 
Two special cases $\set{p=1,\, q=\infty}$ and $\set{p=2,\, q=2}$ are particularly noteworthy, as both reduce to MT-like bounds when the initial state is pure.

{\it Case I: $\set{p=1,\, q=\infty}$.---}
For $p=1$, the Schatten norm corresponds to the trace norm $\norm{A}_1 = \Tr{ |A|}$, which equals the sum of the singular values of $A$ \cite{nielsen_chuang_2010}. 
The Schatten $\infty$-norm $\norm{A}_\infty$ is the largest singular value of $A$. 
Here, we focus on the pure state case, i.e., $\rho_t=\ket{\psi_t}\!\bra{\psi_t}$ with $t\in[0,T]$. 
In this scenario, $\norm{ [H_t,\rho_0] }_\infty = \Delta H_t(\ket{\psi_0}) =\sqrt{\langle \psi_0|H^2_t|\psi_0\rangle-\langle \psi_0|H_t|\psi_0\rangle^2}$, which is precisely the energy uncertainty with respect to the initial state $\ket{\psi_0}$ (Appendix \ref{appsec:Schatten}). 
Additionally, the relative purity $F_{\rm RP}(\rho_0,\rho_T)$ reduces to the quantum fidelity $F(\ket{\psi_0},\ket{\psi_T})=\abs{\braket{\psi_0|\psi_T}}^2$. 
From Eq.~\eqref{eq:Bound_S_pq}, we obtain the following MT-like bound:
\be
\mathcal{T}_{\rm S,\, Pure}(1,\infty) = \frac{1-F(\ket{\psi_0},\ket{\psi_T})}{\frac{1}{T}\int^T_0\d t \, C_1(\ket{\psi_t}) \cdot \Delta H_t(\ket{\psi_0})} \, . 
\label{eq:Bound_S_Pure_p=1}
\ee 
Hereafter, we adopt the notation $C_p(\ket{\psi_t})$ as a substitute for $C_p(\rho_t)$ when $\rho_t=\ket{\psi_t}\!\bra{\psi_t}$ is pure. 
Furthermore, when the Hamiltonian is time-independent, i.e., ${H_t\equiv H}$, we have $C_1(\ket{\psi_t}) \equiv C_1(\ket{\psi_0})$ (Appendix \ref{appsec:Coherence}), and the bound simplifies to
\be
\tilde{\mathcal{T}}_{\rm S,\, Pure}(1,\infty) = \frac{1-F(\ket{\psi_0},\ket{\psi_T})}{C_1(\ket{\psi_0})\cdot \Delta H(\ket{\psi_0})  } \, . 
\label{eq:Bound_S_MT_p=1}
\ee

Comparing our result $\mathcal{T}_{\rm S,\, Pure}(1,\infty)$ to the generalized MT bound [see $T_{\rm AA}$ in Table \ref{tab:compare_QSL}], 
a significant difference is the decomposition of $\Delta H_t(\ket{\psi_t})$ into the product $C_1(\ket{\psi_t}) \cdot \Delta H_t(\ket{\psi_0})$. 
Therefore, our QSLs cleanly separate the contributions of coherence and energy uncertainty, clarifying their distinct roles in quantum dynamics.

{\it Case II: $\set{p=2,\, q=2}$.---}
For $p=2$, the Schatten norm is equivalent to the Hilbert-Schmidt norm.
The coherence measure $C_2(\rho_t)= \min_{\sigma_t \in \mathcal{I}_t}  \norm{\rho_t -  \sigma_t}_2$ is minimized when $\sigma_t=\sum_{n}\braket{n_t|\rho_t|n_t}\ket{n_t}\!\bra{n_t}$ is the diagonal part of $\rho_t$ in the instantaneous energy basis \cite{Plenio2014PRL}. 
Moreover, $\norm{ [H_t,\rho_0] }_2 = \sqrt{2}\sqrt{\Tr{\rho^2_0 H^2_t-(\rho_0 H_t)^2}}$. 
From Eq.~\eqref{eq:Bound_S_pq}, we thus obtain:
\be
\mathcal{T}_{\rm S}(2, 2) 
= \frac{\sbraket{1 - F_{\rm RP}(\rho_0,\rho_T)}\Tr{\rho^2_0}}{\frac{\sqrt{2}}{T}\int^T_0\d t \, C_2(\rho_t) \sqrt{\Tr{\rho^2_0 H^2_t-(\rho_0 H_t)^2}}} \, . 
\label{eq:Bound_S_p=2}
\ee 
For pure states, i.e., $\rho_t=\ket{\psi_t}\!\bra{\psi_t}$ with $t\in[0,T]$, we find $\norm{ [H_t,\rho_0] }_2=\sqrt{2}\, \Delta H_t(\ket{\psi_0})$. 
Here, we used the facts $\rho^2_0=\rho_0$ and $\Tr{(\rho_0 H_t)^2}=\braket{\psi_0|H_t|\psi_0}^2=\Trs{\rho_0 H_t}$. 
This leads to the following MT-like bound:
\be
\mathcal{T}_{\rm S,\, Pure}(2, 2) = \frac{1-F(\ket{\psi_0},\ket{\psi_T})}{\frac{\sqrt{2}}{T}\int^T_0\d t \, C_2(\ket{\psi_t}) \cdot \Delta H_t(\ket{\psi_0})} \, . 
\label{eq:Bound_S_Pure_p=2}
\ee 
Restricting further to the time-independent case, we have ${C_2(\ket{\psi_t}) \equiv C_2(\ket{\psi_0}})$ (Appendix \ref{appsec:Coherence}), reducing the bound to
\be
\tilde{\mathcal{T}}_{\rm S,\, Pure}(2,2) = \frac{1-F(\ket{\psi_0},\ket{\psi_T})}{\sqrt{2}\, C_2(\ket{\psi_0}) \cdot \Delta H(\ket{\psi_0})} \, .
\label{eq:Bound_S_MT_p=2} 
\ee
Once again, these bounds decouple the contributions of coherence and energy uncertainty. 
Compared to the $\set{p=1,q=\infty}$ case, these bounds are generally easier to evaluate. 
The explicit analytical evaluation of $C_2(\ket{\psi_t})$ for a physical model will be provided in Sec. \ref{sec:examples}.

\subsection{Quantum speed limits based on Hellinger distance}
We now derive the second coherent QSL, based on the Hellinger distance $D_{\rm H}(\rho,\sigma) \coloneqq {\rm Tr}[(\sqrt{\rho}-\sqrt{\sigma})^2] = 2- 2F_{\rm A}(\rho,\sigma)$. 
Using the Liouville-von Neumann equation for $\sqrt{\rho_t}$: $\d \sqrt{\rho_t}/\d t = i[\sqrt{\rho_t}, H_t]$ (Appendix \ref{appsec:Liouville}), we bound the rate of change of the quantum affinity $F_{\rm A}(\rho_0,\rho_t)$.
Defining $\tilde{C}_p(\rho_t)= \min_{\sigma_t \in \mathcal{I}_t}  \norm{ \sqrt{\rho_t} -  \sqrt{\sigma_t} }_p$, we obtain (Appendix \ref{appsec:Derivation}):
\be
T \geq \mathcal{T}_{\rm H}(p,q) 
\coloneqq \frac{1-F_{\rm A}(\rho_0,\rho_T)}{\frac{1}{T}\int^T_0\d t \, \tilde{C}_p(\rho_t) \norm{ [H_t,\sqrt{\rho_0}] }_q} \,,
\label{eq:Bound_H_pq}
\ee
which holds for all $p,\,q \in[1,\infty]$ with $1/p+1/q=1$ and constitutes our second main result.

Among this family, the case $\set{p=2,\,q=2}$ is of particular interest. 
For $p=2$, the Schatten norm becomes $\norm{ [H_t,\sqrt{\rho_0}] }_2=\sqrt{2 I(\rho_0,H_t)}$, where $I(\rho_0,H_t)=\frac{1}{2}\Tr{[H_t,\sqrt{\rho_0}]^{\dag}[H_t,\sqrt{\rho_0}]}$ is the WYSI of $\rho_0$ with respect to $H_t$ \cite{Wigner_Yanase_1963}. 
Furthermore, $\tilde{C}_2(\rho_t)$ relates directly to the Hellinger distance: $\tilde{C}_2(\rho_t)= \min_{\sigma_t \in \mathcal{I}_t}  \norm{ \sqrt{\rho_t} -  \sqrt{\sigma_t} }_2 \equiv \sqrt{C_{\rm H}(\rho_t)}$. Here, 
\be
C_{\rm H}(\rho_t) = \min_{\sigma_t \in \mathcal{I}_t}D_{\rm H}(\rho_t,\sigma_t)
\label{eq:C_H}
\ee
is the coherence measure based on Hellinger distance \cite{Jin2018PRA}. 
And the optimal incoherent state achieving the minimum is $\sigma_t = \sum_n \lambda_{n,t}\ket{n_t}\!\bra{n_t}$ with $\lambda_{n,t} = \braket{n_t|\sqrt{\rho_t}|n_t}^2/\sum_{n}\braket{n_t|\sqrt{\rho_t}|n_t}^2$ \cite{Jin2018PRA}. 
From Eq.~\eqref{eq:Bound_H_pq}, we thus obtain 
\be
\mathcal{T}_{\rm H}(2,2) = \frac{1-F_{\rm A}(\rho_0,\rho_T)}{\frac{\sqrt{2}}{T}\int^T_0\d t \,  \sqrt{C_{\rm H}(\rho_t)}\cdot \sqrt{I(\rho_0,H_t)}}  \, .
\label{eq:Bound_H}
\ee
This result provides a geometric alternative to the Schatten bound, connecting QSLs to information geometry. 
Compared to the established bound $T_{\rm WY}$ (Table~\ref{tab:compare_QSL}), our result explicitly disentangles the contribution of coherence from the WYSI of the state, clarifying their respective roles in quantum dynamics. 
Besides, note that while $\mathcal{T}_{\rm S}(2,2)$ and $\mathcal{T}_{\rm H}(2,2)$ share a similar structure, they are generally distinct. For mixed states, the minimizers for $C_2$ and $C_{\rm H}$ are different, leading to distinct physical interpretations.

\begin{figure*}[htp!]
\centering
\includegraphics[width=1\textwidth]{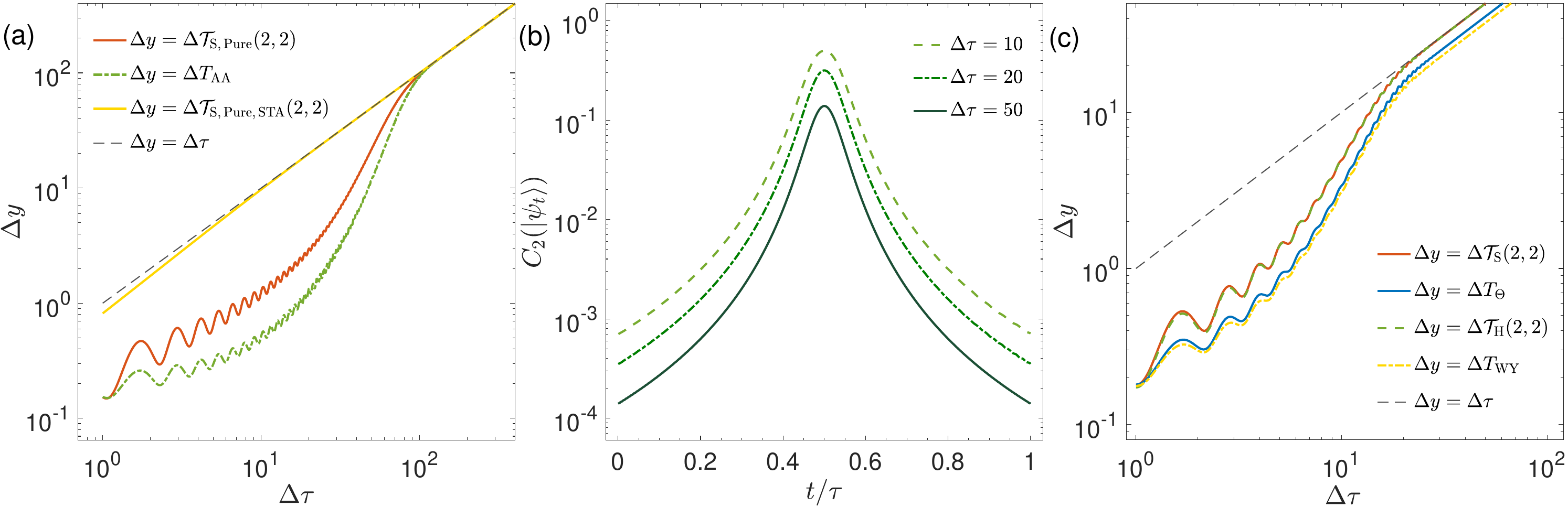}
\caption{
Demonstrations of QSLs for the LZ model and a related two-qubit model. 
(a) The LZ model with pure state. 
The red solid line and the green dashed-dotted line are $\Delta y=\Delta \mathcal{T}_{\rm S,\, Pure}(2,2)$ [Eq.~\eqref{eq:Bound_S_Pure_p=2}] and $\Delta y=\Delta T_{\rm AA}$ (Table~\ref{tab:compare_QSL}), respectively. 
The yellow solid line is the coherent bound $\Delta y=\Delta \mathcal{T}_{\rm S,\, Pure,\, STA}(2,2)$ for the STA dynamics (main text). 
The gray dashed line is the linear reference $\Delta y=\Delta\tau$. 
(b) Coherence measure $C_2(\ket{\psi_{t}})$ during the STA dynamics. We choose a series of adiabatic times, $\Delta\tau=10,\,20,\,50$. 
(c) A two-qubit model with mixed state. 
The red solid line and the green dashed line are $\Delta y=\Delta \mathcal{T}_{\rm S}(2,2)$ [Eq.~\eqref{eq:Bound_S_p=2}] and $\Delta y=\Delta \mathcal{T}_{\rm H}(2,2)$ [Eq.~\eqref{eq:Bound_H}], respectively. 
The blue solid line and the yellow dashed-dotted line are $\Delta y=\Delta T_{\Theta}$ and $\Delta y=\Delta T_{\rm WY}$, respectively (Table~\ref{tab:compare_QSL}). 
The gray dashed line is the linear reference $\Delta y=\Delta\tau$. 
Here, we have chosen the linear ramp $J_t=J(-1+2t/\tau)$ with $J=10\Delta$. 
The initial states in (a, b) and (c) are set to be $\ket{\psi_0}$ and $\varrho_0(\eta=3/5)$, respectively. 
And the dimensionless times are scaled by the energy splitting $\Delta$. 
}
\label{fig:Fig_1}
\end{figure*}


\section{Examples}
\label{sec:examples}
We now analyze our coherent QSLs and compare them with the established results summarized in Table~\ref{tab:compare_QSL}.
As a concrete example, we consider the paradigmatic Landau-Zener (LZ) model, which describes a time-dependent two-level crossing and has broad applications in quantum physics \cite{Ivakhnenko_2023}.
In its most basic form, the Hamiltonian is given by
\be
H_{\rm LZ}(t) = \Delta\sigma^x+J_t\sigma^z \,,
\ee
where $\sigma^{x,y,z}$ are the standard Pauli matrices, $\Delta$ is the energy splitting, and the linear ramp $J_t=J(-1+2t/\tau)$ represents the time-dependent field with $\tau$ being the total evolution time. 
Hereafter, we set $\Delta$ as the energy unit. 
Without loss of generality, we present results for the $\set{p=2,q=2}$ case. 

First, we consider the pure state case where the initial state is the ground state of $H_{\rm LZ}(0)$, denoted as $\ket{\psi_0}$. 
To explicitly evaluate the coherent bound, we work in the instantaneous energy eigenbasis $\set{\ket{n_t}}$.
The instantaneous eigenstates of $H_{\rm LZ}(t)$ are given by $\ket{0_t} = \cos(\theta_t/2) \ket{0} + \sin(\theta_t/2) \ket{1}$ and $\ket{1_t} = -\sin(\theta_t/2) \ket{0} + \cos(\theta_t/2) \ket{1}$, where the mixing angle $\theta_t$ is defined by $\tan\theta_t = \Delta/J_t$ with $\theta_t \in [0, \pi]$ and $\{\ket{0}, \ket{1}\}$ is the computational basis ($\sigma^z \ket{0} = \ket{0}, \sigma^z \ket{1} = -\ket{1}$). 
This basis $\{\ket{0_t}, \ket{1_t}\}$ defines the reference set of incoherent states $\mathcal{I}_t$ at each time instant $t$. 
Let the time-evolved state be $\rho_t = \ket{\psi_t}\!\bra{\psi_t}$ with $\ket{\psi_t} = c_0(t) \ket{0_t} + c_1(t) \ket{1_t}$.
For the Schatten 2-norm case, the coherence measure $C_2(\rho_t) = \min_{\sigma_t \in \mathcal{I}_t} \norm{\rho_t - \sigma_t}_2$ is obtained by choosing $\sigma_t$ as the diagonal part of $\rho_t$ in the basis $\{\ket{0_t}, \ket{1_t}\}$.
Explicitly, we have $\sigma_t = |c_0(t)|^2 \ket{0_t}\!\bra{0_t} + |c_1(t)|^2 \ket{1_t}\!\bra{1_t}$.
The coherence measure is then calculated as $C_2(\rho_t)  = \norm{\rho_t - \sigma_t}_2 = \sqrt{2 P_{\rm exc}(t) [1 - P_{\rm exc}(t)]}$, where $P_{\rm exc}(t) = |c_1(t)|^2$ is the instantaneous excitation probability.
This result clearly shows that coherence in the energy basis is directly proportional to the mixing of energy levels.
The energy uncertainty term in the denominator of our bound is given by $\norm{[H_t, \rho_0]}_2 = \sqrt{2}\Delta H_t(\ket{\psi_0})$ (see Appendix \ref{appsec:Schatten}).
For the LZ model with $\ket{\psi_0}=\ket{1_{t=0}}$, we have $\Delta H_t(\ket{\psi_0}) = \sqrt{\Omega_t^2 - (J_0 J_t + \Delta^2)^2/\Omega_0^2}$, where $\Omega_t = \sqrt{\Delta^2 + J_t^2}$.
Substituting these explicit expressions into Eq.~\eqref{eq:Bound_S_Pure_p=2}, we obtain the coherent QSL for the LZ model.

Numerical results are displayed in Fig.~\ref{fig:Fig_1} (a). We observe that our coherent bound $\mathcal{T}_{\rm S,\, Pure}(2,2)$ is tighter than the corresponding established bound $T_{\rm AA}$ (Table~\ref{tab:compare_QSL}). 
Crucially, this example demonstrates that our QSLs are asymptotically saturated in the adiabatic limit ($\tau \to \infty$).
The underlying mechanism for this saturation stems from the system state $\ket{\psi_t}$ strictly following the instantaneous ground state of $H_{\rm LZ}(t)$ as the adiabatic time increases.
To illustrate this, we examine the saturation of H\"older's inequality [Eq.~\eqref{eq:Holder_rho}] at the critical moment $t_c=\tau/2$, where the energy gap is minimal and the adiabatic condition is most stringent.
Defining the state $\ket{\psi_{t_c}}\coloneqq\ket{\psi_{c}}=(a_{c},b_{c})^T$, the corresponding $\varrho_{c} := \rho_c - \sigma_c =\ket{\psi_{c}}\!\bra{\psi_{c}} - \sigma_{c}$ can be expressed in the computational basis as
\be
\varrho_{c} =\frac{1}{2}\left(\begin{array}{cc}
D_{c} & O_{c} \\
-O_{c} & -D_{c}
\end{array}\right) \,,
\ee
where $D_{c}=\abs{a_{c}}^2-\abs{b_{c}}^2$ and $O_{c}=a_{c}b^\ast_{c}-a^\ast_{c}b_{c}$.
In the adiabatic limit, $\ket{\psi_{c}}$ approaches the ground state of $H_{\rm LZ}(t_c) = \Delta\sigma^x$, implying $D_{c}\approx 0$.
Consequently, the diagonal elements of $\varrho_{c}$ vanish, making $\varrho_{c}$ approximately an antisymmetric $2\times 2$ matrix.
Since both $H_{\rm LZ}(t_c)$ and $\ket{\psi_0}\!\bra{\psi_0}$ are
real matrices, $A_{c}=-i[H_{\rm LZ}(t_c),\ket{\psi_0}\!\bra{\psi_0}]$ is also antisymmetric.
Therefore, $\varrho_{c}\approx\lambda_{c} A_{c}$ for some constant $\lambda_{c}$.
This proportionality ensures the saturation of H\"older's inequality $\Tr{ \varrho_{c} A_{c}} \leq C_2(\ket{\psi_{c}}) \norm{ A_{c} }_2$, as the inequality becomes an equality if and only if one matrix is a scalar multiple of the other. This confirms the asymptotic saturation of our coherent QSLs.

A natural question arises: can coherent QSLs be saturated at finite time?
We find that this is achievable using shortcuts to adiabaticity (STA) techniques \cite{Muga_2019_RMP}.
For the LZ Hamiltonian $H_{\rm LZ}(t)$, we can construct a counter-diabatic Hamiltonian $H_{\rm STA}(t)=H_{\rm LZ}(t)+V(t)$ such that the adiabatic solution of $H_{\rm LZ}(t)$ becomes an exact solution of the
Schr\"odinger equation governed by $H_{\rm STA}(t)$. The auxiliary potential is given by \cite{Berry_2009}:
\be
V(t) = -\frac{\d J_t}{\d t}\frac{\Delta}{2(\Delta^2+J^2_t)}\sigma^y \,.
\ee
Using the same initial state $\ket{\psi_0}$, we plot the corresponding coherent bound $\mathcal{T}_{\rm S,\, Pure,\,STA}(2,2)$ in Fig.~\ref{fig:Fig_1} (a) (yellow solid line).
It is evident that the STA bound is saturated much earlier than the standard adiabatic case.
More importantly, this finite-time saturation reveals the fundamental role of coherence as a resource.
In Fig.~\ref{fig:Fig_1} (b), we plot the coherence measure versus evolution time.
We observe that as the evolution time decreases (i.e., faster dynamics), the required coherence $C_2(\ket{\psi_{t}})$ increases significantly. 
This clearly shows that generating faster quantum evolution is fueled by a greater consumption of coherence, thereby establishing coherence as a key kinematic resource for quantum speed.

For the mixed state case, we consider a related two-qubit composite system with the Hamiltonian:
\be
H_{\rm TQ}(t) = \Delta(\sigma^x_1+\sigma^x_2)+J_t\sigma^z_1\sigma^z_2\, .
\ee
The initial state is chosen as $\varrho_0(\eta)=\eta\ket{\varphi_0}\!\bra{\varphi_0}+(1-\eta)\ket{\varphi_1}\!\bra{\varphi_1}$, where $\ket{\varphi_0}$ and $\ket{\varphi_1}$ are the ground and first excited states of $H_{\rm TQ}(0)$, respectively. 
As shown in Fig.~\ref{fig:Fig_1} (c), both $\mathcal{T}_{\rm S}(2,2)$ [Eq.~\eqref{eq:Bound_S_p=2}] and $\mathcal{T}_{\rm H}(2,2)$ [Eq.~\eqref{eq:Bound_H}] are tighter than the established bounds $T_{\Theta}$ and $T_{\rm WY}$ (Table~\ref{tab:compare_QSL}). 
We have verified that the saturation of our coherent QSLs in the adiabatic limit holds here as well. In contrast, the established bounds $T_{\Theta}$ and $T_{\rm WY}$ fail to saturate in this regime.


\section{CONCLUDING REMARKS} 
\label{sec:conclusion}
We have established a unified theoretical framework for coherent QSLs in general unitary dynamics, measuring coherence via Schatten $p$-norms and the Hellinger distance. 
The core of our approach lies in introducing an instantaneous incoherent reference state in the energy eigenbasis and employing relative purity and quantum affinity to quantify state divergence. 
The resulting coherent QSLs are tighter than established bounds for the models investigated and are asymptotically saturable in the adiabatic limit.
Furthermore, using STA techniques, we demonstrated that these bounds can be saturated at finite time. 
Crucially, the analysis of STA dynamics reinforces the interpretation of coherence as a critical kinematic resource: the speed of state evolution is directly correlated with the availability of coherence in the system.
Our findings highlight the central role of coherence in quantum dynamics and are expected to find broad applications in quantum control.
The results from the LZ and two-qubit models suggest potential extensions to many-body systems such as the quantum Ising model \cite{delCampo_2012_PRL}, with relevance for quantum simulation platforms including superconducting qubits \cite{Johnson_2011_Nature}, Rydberg atoms \cite{Scholl_2021_Nature}, and trapped ions \cite{Duan_2023_PRXQuantum}.
Future work may extend these concepts to non-unitary dynamics involving decoherence \cite{Deffner2013PRL, delCampo2013PRL, Taddei2013PRL, Marvian2015PRL}.

\begin{acknowledgments}

We are grateful to the anonymous Referees for their constructive comments. 
H.W. is supported by the National Natural Science Foundation of China (12401158) and the Fundamental Research Funds for the Central Universities (30925010422). 
X.Q. acknowledges support from the Shanghai Science and Technology project (24LZ1401600) and the Fundamental Research Funds for the Central Universities (22120240278).

\end{acknowledgments}

\section*{DATA AVAILABILITY}

The data that support the findings of this study are available from the authors upon reasonable request.

\appendix

\section{Derivation of the Coherent QSLs}
\label{appsec:Derivation}
Here we provide the detailed derivations for the two families of coherent QSLs.
We start with the Liouville-von Neumann equation $\d\rho_t/\d t=i[\rho_t,H_t]$.
The rate of change of relative purity is given by:
\be
-\frac{\d F_{\rm RP}(\rho_0,\rho_t)}{\d t} = \frac{-i\Tr{\rho_0[\rho_t,H_t]}}{\Tr{\rho^2_0}} \, .
\label{eq:dF_RP_app}
\ee
Let $\sigma_t$ be any incoherent state in the instantaneous set $\mathcal{I}_t$, such that $[\sigma_t, H_t]=0$. We can rewrite the commutator as $[\rho_t,H_t]=[\rho_t-\sigma_t,H_t]$.
The numerator of Eq.~\eqref{eq:dF_RP_app} can be bounded using H\"older's inequality:
\bea
-i\Tr{\rho_0[\rho_t,H_t]} 
&=& -i\Tr{\rho_0[\rho_t-\sigma_t,H_t]} \nn \\
&=& -i\Tr{(\rho_t-\sigma_t)[H_t,\rho_0]} \nn \\
&\leq& \norm{\rho_t-\sigma_t}_p \norm{[H_t,\rho_0]}_q \, ,
\label{eq:Holder_rho}
\eea
where $1/p+1/q=1$ \cite{Bhatia2013matrix}. Since this holds for any $\sigma_t \in \mathcal{I}_t$, we take the minimum over all such states to tighten the bound.
Using the definition $C_p(\rho_t)= \min_{\sigma_t \in \mathcal{I}_t} \norm{\rho_t -  \sigma_t}_p$, we obtain the fundamental differential inequality:
\be
-\frac{\d F_{\rm RP}(\rho_0,\rho_t)}{\d t} \leq \frac{C_p(\rho_t) \norm{[H_t,\rho_0]}_q}{\Tr{\rho^2_0}} \, .
\ee
Integrating from $t=0$ to $t=T$ directly yields Eq.~\eqref{eq:Bound_S_pq} in the main text.

Similarly, for the Hellinger distance family, we consider the Liouville equation for the square root of the density matrix: $\d \sqrt{\rho_t}/\d t = i[\sqrt{\rho_t}, H_t]$ (Appendix \ref{appsec:Liouville}).
The rate of change of quantum affinity $F_{\rm A}(\rho_0,\rho_t)={\rm Tr}(\sqrt{\rho_0}\sqrt{\rho_t})$ is: 
\be
-\frac{\d F_{\rm A}(\rho_0,\rho_t)}{\d t} = -i \Tr{\sqrt{\rho_0}[\sqrt{\rho_t}, H_t] }\, .
\ee
Introducing $\sqrt{\sigma_t}$ where $\sigma_t \in \mathcal{I}_t$ (so $[\sqrt{\sigma_t}, H_t]=0$), we have $[\sqrt{\rho_t}, H_t] = [\sqrt{\rho_t}-\sqrt{\sigma_t}, H_t]$.
Applying H\"older's inequality:
\bea
&& -i\Tr{\sqrt{\rho_0}[\sqrt{\rho_t},H_t]} \nn \\
&\leq& \norm{\sqrt{\rho_t}-\sqrt{\sigma_t}}_p \norm{[H_t,\sqrt{\rho_0}]}_q \, .
\eea
Minimizing over $\sigma_t$ yields $\tilde{C}_p(\rho_t) \norm{[H_t,\sqrt{\rho_0}]}_q$, which upon integration gives Eq.~\eqref{eq:Bound_H_pq}.

\section{Schatten norm $\norm{[H_t,\rho_0]}_p$ for pure state}
\label{appsec:Schatten}
Here we derive the relationship between the Schatten norm $\norm{[H_t,\rho_0]}_p$ and the energy uncertainty for a pure initial state $\rho_0=\ket{\psi_0}\!\bra{\psi_0}$ and arbitrary $p \in [1, \infty]$.
Let the Hamiltonian acting on the state be decomposed as:
\be
H_t\ket{\psi_0}=\bar{H}_t\ket{\psi_0}+\Delta H_t(\ket{\psi_0})\ket{\psi_0^\perp}\, ,
\ee
where $\bar{H}_t = \braket{\psi_0|H_t|\psi_0}$ is the mean energy, $\Delta H_t(\ket{\psi_0}) = \sqrt{\langle \psi_0|H^2_t|\psi_0\rangle-\langle \psi_0|H_t|\psi_0\rangle^2}$ is the energy uncertainty, and $\ket{\psi_0^\perp}$ is a normalized state orthogonal to $\ket{\psi_0}$.
The commutator $G_t \coloneqq [H_t, \rho_0]$ can be written as:
\bea
G_t &=& H_t\ket{\psi_0}\!\bra{\psi_0} - \ket{\psi_0}\!\bra{\psi_0}H_t \nn \\
&=& \Delta H_t(\ket{\psi_0}) \left( \ket{\psi_0^\perp}\!\bra{\psi_0} - \ket{\psi_0}\!\bra{\psi_0^\perp} \right) \, .
\eea
The matrix $G_t^\dagger G_t$ is thus diagonal in the subspace $\{\ket{\psi_0}, \ket{\psi_0^\perp}\}$:
\be
G_t^\dagger G_t = [\Delta H_t(\ket{\psi_0})]^2 \left(\ket{\psi_0}\!\bra{\psi_0}+\ket{\psi_0^\perp}\!\bra{\psi_0^\perp}\right) \, .
\ee
The singular values of $G_t$ are the square roots of the eigenvalues of $G_t^\dagger G_t$. Thus, $G_t$ has two non-zero singular values, both equal to $\Delta H_t(\ket{\psi_0})$. 
Therefore, we have: 
\bea
\norm{[H_t,\rho_0]}_p 
&=& \left( [\Delta H_t(\ket{\psi_0})]^p + [\Delta H_t(\ket{\psi_0})]^p \right)^{1/p} \nn \\
&=& 2^{1/p} \Delta H_t(\ket{\psi_0}) \, .
\eea
This result generalizes the relations used in the main text:
\begin{itemize}
\item For $p=\infty$ (Schatten $\infty$-norm), $2^{1/\infty}=1$, yielding $\norm{[H_t,\rho_0]}_\infty = \Delta H_t(\ket{\psi_0})$.
\item For $p=2$ (Hilbert-Schmidt norm), $2^{1/2}=\sqrt{2}$, yielding $\norm{[H_t,\rho_0]}_2 = \sqrt{2} \Delta H_t(\ket{\psi_0})$.
\end{itemize}

\section{Coherence measure for time-independent Hamiltonian}
\label{appsec:Coherence}
Here we show that for a time-independent Hamiltonian $H_t\equiv H$, the coherence measure $C_p(\rho_t)$ is time-independent.
The norm $\norm{A}_p$ is determined solely by the singular values of $A$ and is therefore invariant under unitary transformations.
Thus, for a time-independent Hamiltonian $H$, we have
\bea
\norm{\rho_t-\sigma}_p
&=& \norm{ e^{-iHt}\rho_0 e^{iHt}-e^{-iHt}\sigma e^{iHt}}_p \nn \\
&=& \norm{\rho_0 - \sigma }_p\, ,
\eea
where $\sigma$ is an incoherent state such that $[H,\sigma]=0$. Since $\sigma$ commutes with $H$, the unitary rotation leaves the set of incoherent states invariant.
Consequently, $\min_{\sigma} \norm{\rho_t-\sigma}_p=\min_{\sigma}\norm{\rho_0 - \sigma }_p$, implying $C_p(\rho_t) \equiv C_p(\rho_0)$ for all $t\geq 0$ and $p\geq 1$.

\section{Liouville-von Neumann equation for $\sqrt{\rho_t}$}
\label{appsec:Liouville}
Here we show that $\sqrt{\rho_t}$ also satisfies the Liouville-von Neumann equation.
Considering the spectral decomposition of the initial state $\rho_0=\sum_{n}\lambda_n\ket{u^n_0}\!\bra{u^n_0}$, we thus have the time-evolved state
\be
\rho_t=U_t\rho_0U^\dag_t=\sum_{n} \lambda_n\ket{u^n_t}\!\bra{u^n_t}\, ,
\ee
where $\ket{u^n_t}=U_t\ket{u^n_0}$ satisfies the Schr\"odinger equation $i\d \ket{u^n_t}/\d t = H_t\ket{u^n_t}$. Therefore, we have $\sqrt{\rho_t}=\sum_{n}\sqrt{\lambda_n}\ket{u^n_t}\!\bra{u^n_t}$, which directly implies the Liouville-von Neumann equation: $\d \sqrt{\rho_t}/\d t = i[\sqrt{\rho_t}, H_t]$.

\bibliography{References}

\begin{thebibliography}{68}%
\makeatletter
\providecommand \@ifxundefined [1]{%
 \@ifx{#1\undefined}
}%
\providecommand \@ifnum [1]{%
 \ifnum #1\expandafter \@firstoftwo
 \else \expandafter \@secondoftwo
 \fi
}%
\providecommand \@ifx [1]{%
 \ifx #1\expandafter \@firstoftwo
 \else \expandafter \@secondoftwo
 \fi
}%
\providecommand \natexlab [1]{#1}%
\providecommand \enquote  [1]{``#1''}%
\providecommand \bibnamefont  [1]{#1}%
\providecommand \bibfnamefont [1]{#1}%
\providecommand \citenamefont [1]{#1}%
\providecommand \href@noop [0]{\@secondoftwo}%
\providecommand \href [0]{\begingroup \@sanitize@url \@href}%
\providecommand \@href[1]{\@@startlink{#1}\@@href}%
\providecommand \@@href[1]{\endgroup#1\@@endlink}%
\providecommand \@sanitize@url [0]{\catcode `\\12\catcode `\$12\catcode
  `\&12\catcode `\#12\catcode `\^12\catcode `\_12\catcode `\%12\relax}%
\providecommand \@@startlink[1]{}%
\providecommand \@@endlink[0]{}%
\providecommand \url  [0]{\begingroup\@sanitize@url \@url }%
\providecommand \@url [1]{\endgroup\@href {#1}{\urlprefix }}%
\providecommand \urlprefix  [0]{URL }%
\providecommand \Eprint [0]{\href }%
\providecommand \doibase [0]{https://doi.org/}%
\providecommand \selectlanguage [0]{\@gobble}%
\providecommand \bibinfo  [0]{\@secondoftwo}%
\providecommand \bibfield  [0]{\@secondoftwo}%
\providecommand \translation [1]{[#1]}%
\providecommand \BibitemOpen [0]{}%
\providecommand \bibitemStop [0]{}%
\providecommand \bibitemNoStop [0]{.\EOS\space}%
\providecommand \EOS [0]{\spacefactor3000\relax}%
\providecommand \BibitemShut  [1]{\csname bibitem#1\endcsname}%
\let\auto@bib@innerbib\@empty
\bibitem [{\citenamefont {Frey}(2016)}]{Frey2016}%
  \BibitemOpen
  \bibfield  {author} {\bibinfo {author} {\bibfnamefont {M.~R.}\ \bibnamefont
  {Frey}},\ }\bibfield  {title} {\bibinfo {title} {{Quantum speed
  limits---primer, perspectives, and potential future directions}},\ }\href
  {https://doi.org/10.1007/s11128-016-1405-x} {\bibfield  {journal} {\bibinfo
  {journal} {Quantum Information Processing}\ }\textbf {\bibinfo {volume}
  {15}},\ \bibinfo {pages} {3919} (\bibinfo {year} {2016})}\BibitemShut
  {NoStop}%
\bibitem [{\citenamefont {Deffner}\ and\ \citenamefont
  {Campbell}(2017)}]{Deffner_2017}%
  \BibitemOpen
  \bibfield  {author} {\bibinfo {author} {\bibfnamefont {S.}~\bibnamefont
  {Deffner}}\ and\ \bibinfo {author} {\bibfnamefont {S.}~\bibnamefont
  {Campbell}},\ }\bibfield  {title} {\bibinfo {title} {{Quantum speed limits:
  from Heisenberg’s uncertainty principle to optimal quantum control}},\
  }\href {https://doi.org/10.1088/1751-8121/aa86c6} {\bibfield  {journal}
  {\bibinfo  {journal} {Journal of Physics A: Mathematical and Theoretical}\
  }\textbf {\bibinfo {volume} {50}},\ \bibinfo {pages} {453001} (\bibinfo
  {year} {2017})}\BibitemShut {NoStop}%
\bibitem [{\citenamefont {Bekenstein}(1981)}]{Bekenstein1981PRL}%
  \BibitemOpen
  \bibfield  {author} {\bibinfo {author} {\bibfnamefont {J.~D.}\ \bibnamefont
  {Bekenstein}},\ }\bibfield  {title} {\bibinfo {title} {{Energy Cost of
  Information Transfer}},\ }\href {https://doi.org/10.1103/PhysRevLett.46.623}
  {\bibfield  {journal} {\bibinfo  {journal} {Phys. Rev. Lett.}\ }\textbf
  {\bibinfo {volume} {46}},\ \bibinfo {pages} {623} (\bibinfo {year}
  {1981})}\BibitemShut {NoStop}%
\bibitem [{\citenamefont {Deffner}(2020)}]{Deffner2020PRR}%
  \BibitemOpen
  \bibfield  {author} {\bibinfo {author} {\bibfnamefont {S.}~\bibnamefont
  {Deffner}},\ }\bibfield  {title} {\bibinfo {title} {Quantum speed limits and
  the maximal rate of information production},\ }\href
  {https://doi.org/10.1103/PhysRevResearch.2.013161} {\bibfield  {journal}
  {\bibinfo  {journal} {Phys. Rev. Res.}\ }\textbf {\bibinfo {volume} {2}},\
  \bibinfo {pages} {013161} (\bibinfo {year} {2020})}\BibitemShut {NoStop}%
\bibitem [{\citenamefont {Lloyd}(2000)}]{Lloyd2000}%
  \BibitemOpen
  \bibfield  {author} {\bibinfo {author} {\bibfnamefont {S.}~\bibnamefont
  {Lloyd}},\ }\bibfield  {title} {\bibinfo {title} {Ultimate physical limits to
  computation},\ }\href {https://doi.org/10.1038/35023282} {\bibfield
  {journal} {\bibinfo  {journal} {Nature}\ }\textbf {\bibinfo {volume} {406}},\
  \bibinfo {pages} {1047} (\bibinfo {year} {2000})}\BibitemShut {NoStop}%
\bibitem [{\citenamefont {Deffner}(2021)}]{Deffner_2021}%
  \BibitemOpen
  \bibfield  {author} {\bibinfo {author} {\bibfnamefont {S.}~\bibnamefont
  {Deffner}},\ }\bibfield  {title} {\bibinfo {title} {{Energetic cost of
  Hamiltonian quantum gates}},\ }\href
  {https://doi.org/10.1209/0295-5075/134/40002} {\bibfield  {journal} {\bibinfo
   {journal} {Europhysics Letters}\ }\textbf {\bibinfo {volume} {134}},\
  \bibinfo {pages} {40002} (\bibinfo {year} {2021})}\BibitemShut {NoStop}%
\bibitem [{\citenamefont {Giovannetti}\ \emph {et~al.}(2011)\citenamefont
  {Giovannetti}, \citenamefont {Lloyd},\ and\ \citenamefont
  {Maccone}}]{Giovannetti2011NatPhoton}%
  \BibitemOpen
  \bibfield  {author} {\bibinfo {author} {\bibfnamefont {V.}~\bibnamefont
  {Giovannetti}}, \bibinfo {author} {\bibfnamefont {S.}~\bibnamefont {Lloyd}},\
  and\ \bibinfo {author} {\bibfnamefont {L.}~\bibnamefont {Maccone}},\
  }\bibfield  {title} {\bibinfo {title} {Advances in quantum metrology},\
  }\href {https://doi.org/10.1038/nphoton.2011.35} {\bibfield  {journal}
  {\bibinfo  {journal} {Nature Photonics}\ }\textbf {\bibinfo {volume} {5}},\
  \bibinfo {pages} {222} (\bibinfo {year} {2011})}\BibitemShut {NoStop}%
\bibitem [{\citenamefont {Beau}\ and\ \citenamefont {del
  Campo}(2017)}]{delCampo2017PRL}%
  \BibitemOpen
  \bibfield  {author} {\bibinfo {author} {\bibfnamefont {M.}~\bibnamefont
  {Beau}}\ and\ \bibinfo {author} {\bibfnamefont {A.}~\bibnamefont {del
  Campo}},\ }\bibfield  {title} {\bibinfo {title} {{Nonlinear Quantum Metrology
  of Many-Body Open Systems}},\ }\href
  {https://doi.org/10.1103/PhysRevLett.119.010403} {\bibfield  {journal}
  {\bibinfo  {journal} {Phys. Rev. Lett.}\ }\textbf {\bibinfo {volume} {119}},\
  \bibinfo {pages} {010403} (\bibinfo {year} {2017})}\BibitemShut {NoStop}%
\bibitem [{\citenamefont {Caneva}\ \emph {et~al.}(2009)\citenamefont {Caneva},
  \citenamefont {Murphy}, \citenamefont {Calarco}, \citenamefont {Fazio},
  \citenamefont {Montangero}, \citenamefont {Giovannetti},\ and\ \citenamefont
  {Santoro}}]{Caneva2009PRL}%
  \BibitemOpen
  \bibfield  {author} {\bibinfo {author} {\bibfnamefont {T.}~\bibnamefont
  {Caneva}}, \bibinfo {author} {\bibfnamefont {M.}~\bibnamefont {Murphy}},
  \bibinfo {author} {\bibfnamefont {T.}~\bibnamefont {Calarco}}, \bibinfo
  {author} {\bibfnamefont {R.}~\bibnamefont {Fazio}}, \bibinfo {author}
  {\bibfnamefont {S.}~\bibnamefont {Montangero}}, \bibinfo {author}
  {\bibfnamefont {V.}~\bibnamefont {Giovannetti}},\ and\ \bibinfo {author}
  {\bibfnamefont {G.~E.}\ \bibnamefont {Santoro}},\ }\bibfield  {title}
  {\bibinfo {title} {{Optimal Control at the Quantum Speed Limit}},\ }\href
  {https://doi.org/10.1103/PhysRevLett.103.240501} {\bibfield  {journal}
  {\bibinfo  {journal} {Phys. Rev. Lett.}\ }\textbf {\bibinfo {volume} {103}},\
  \bibinfo {pages} {240501} (\bibinfo {year} {2009})}\BibitemShut {NoStop}%
\bibitem [{\citenamefont {Brody}\ and\ \citenamefont
  {Meier}(2015)}]{Brody2015PRL}%
  \BibitemOpen
  \bibfield  {author} {\bibinfo {author} {\bibfnamefont {D.~C.}\ \bibnamefont
  {Brody}}\ and\ \bibinfo {author} {\bibfnamefont {D.~M.}\ \bibnamefont
  {Meier}},\ }\bibfield  {title} {\bibinfo {title} {{Solution to the Quantum
  Zermelo Navigation Problem}},\ }\href
  {https://doi.org/10.1103/PhysRevLett.114.100502} {\bibfield  {journal}
  {\bibinfo  {journal} {Phys. Rev. Lett.}\ }\textbf {\bibinfo {volume} {114}},\
  \bibinfo {pages} {100502} (\bibinfo {year} {2015})}\BibitemShut {NoStop}%
\bibitem [{\citenamefont {Campbell}\ and\ \citenamefont
  {Deffner}(2017)}]{Campbell2017PRL}%
  \BibitemOpen
  \bibfield  {author} {\bibinfo {author} {\bibfnamefont {S.}~\bibnamefont
  {Campbell}}\ and\ \bibinfo {author} {\bibfnamefont {S.}~\bibnamefont
  {Deffner}},\ }\bibfield  {title} {\bibinfo {title} {{Trade-Off Between Speed
  and Cost in Shortcuts to Adiabaticity}},\ }\href
  {https://doi.org/10.1103/PhysRevLett.118.100601} {\bibfield  {journal}
  {\bibinfo  {journal} {Phys. Rev. Lett.}\ }\textbf {\bibinfo {volume} {118}},\
  \bibinfo {pages} {100601} (\bibinfo {year} {2017})}\BibitemShut {NoStop}%
\bibitem [{\citenamefont {Campaioli}\ \emph {et~al.}(2022)\citenamefont
  {Campaioli}, \citenamefont {shui Yu}, \citenamefont {Pollock},\ and\
  \citenamefont {Modi}}]{Campaioli_2022_NJP}%
  \BibitemOpen
  \bibfield  {author} {\bibinfo {author} {\bibfnamefont {F.}~\bibnamefont
  {Campaioli}}, \bibinfo {author} {\bibfnamefont {C.}~\bibnamefont {shui Yu}},
  \bibinfo {author} {\bibfnamefont {F.~A.}\ \bibnamefont {Pollock}},\ and\
  \bibinfo {author} {\bibfnamefont {K.}~\bibnamefont {Modi}},\ }\bibfield
  {title} {\bibinfo {title} {Resource speed limits: maximal rate of resource
  variation},\ }\href {https://doi.org/10.1088/1367-2630/ac7346} {\bibfield
  {journal} {\bibinfo  {journal} {New Journal of Physics}\ }\textbf {\bibinfo
  {volume} {24}},\ \bibinfo {pages} {065001} (\bibinfo {year}
  {2022})}\BibitemShut {NoStop}%
\bibitem [{\citenamefont {Mohan}\ \emph {et~al.}(2022)\citenamefont {Mohan},
  \citenamefont {Das},\ and\ \citenamefont {Pati}}]{Mohan_2022_NJP}%
  \BibitemOpen
  \bibfield  {author} {\bibinfo {author} {\bibfnamefont {B.}~\bibnamefont
  {Mohan}}, \bibinfo {author} {\bibfnamefont {S.}~\bibnamefont {Das}},\ and\
  \bibinfo {author} {\bibfnamefont {A.~K.}\ \bibnamefont {Pati}},\ }\bibfield
  {title} {\bibinfo {title} {{Quantum speed limits for information and
  coherence}},\ }\href {https://doi.org/10.1088/1367-2630/ac753c} {\bibfield
  {journal} {\bibinfo  {journal} {New Journal of Physics}\ }\textbf {\bibinfo
  {volume} {24}},\ \bibinfo {pages} {065003} (\bibinfo {year}
  {2022})}\BibitemShut {NoStop}%
\bibitem [{\citenamefont {Deffner}\ and\ \citenamefont
  {Lutz}(2010)}]{Deffner2010PRL}%
  \BibitemOpen
  \bibfield  {author} {\bibinfo {author} {\bibfnamefont {S.}~\bibnamefont
  {Deffner}}\ and\ \bibinfo {author} {\bibfnamefont {E.}~\bibnamefont {Lutz}},\
  }\bibfield  {title} {\bibinfo {title} {{Generalized Clausius Inequality for
  Nonequilibrium Quantum Processes}},\ }\href
  {https://doi.org/10.1103/PhysRevLett.105.170402} {\bibfield  {journal}
  {\bibinfo  {journal} {Phys. Rev. Lett.}\ }\textbf {\bibinfo {volume} {105}},\
  \bibinfo {pages} {170402} (\bibinfo {year} {2010})}\BibitemShut {NoStop}%
\bibitem [{\citenamefont {Campaioli}\ \emph {et~al.}(2017)\citenamefont
  {Campaioli}, \citenamefont {Pollock}, \citenamefont {Binder}, \citenamefont
  {C\'eleri}, \citenamefont {Goold}, \citenamefont {Vinjanampathy},\ and\
  \citenamefont {Modi}}]{Campaioli2017PRL}%
  \BibitemOpen
  \bibfield  {author} {\bibinfo {author} {\bibfnamefont {F.}~\bibnamefont
  {Campaioli}}, \bibinfo {author} {\bibfnamefont {F.~A.}\ \bibnamefont
  {Pollock}}, \bibinfo {author} {\bibfnamefont {F.~C.}\ \bibnamefont {Binder}},
  \bibinfo {author} {\bibfnamefont {L.}~\bibnamefont {C\'eleri}}, \bibinfo
  {author} {\bibfnamefont {J.}~\bibnamefont {Goold}}, \bibinfo {author}
  {\bibfnamefont {S.}~\bibnamefont {Vinjanampathy}},\ and\ \bibinfo {author}
  {\bibfnamefont {K.}~\bibnamefont {Modi}},\ }\bibfield  {title} {\bibinfo
  {title} {{Enhancing the Charging Power of Quantum Batteries}},\ }\href
  {https://doi.org/10.1103/PhysRevLett.118.150601} {\bibfield  {journal}
  {\bibinfo  {journal} {Phys. Rev. Lett.}\ }\textbf {\bibinfo {volume} {118}},\
  \bibinfo {pages} {150601} (\bibinfo {year} {2017})}\BibitemShut {NoStop}%
\bibitem [{\citenamefont {Hasegawa}(2023)}]{Hasegawa2023NC}%
  \BibitemOpen
  \bibfield  {author} {\bibinfo {author} {\bibfnamefont {Y.}~\bibnamefont
  {Hasegawa}},\ }\bibfield  {title} {\bibinfo {title} {{Unifying speed limit,
  thermodynamic uncertainty relation and Heisenberg principle via bulk-boundary
  correspondence}},\ }\href {https://doi.org/10.1038/s41467-023-38074-8}
  {\bibfield  {journal} {\bibinfo  {journal} {Nature Communications}\ }\textbf
  {\bibinfo {volume} {14}},\ \bibinfo {pages} {2828} (\bibinfo {year}
  {2023})}\BibitemShut {NoStop}%
\bibitem [{\citenamefont {Gyhm}\ \emph {et~al.}(2024)\citenamefont {Gyhm},
  \citenamefont {Rosa},\ and\ \citenamefont {\ifmmode~\check{S}\else
  \v{S}\fi{}afr\'anek}}]{Gyhm_2024_PRA}%
  \BibitemOpen
  \bibfield  {author} {\bibinfo {author} {\bibfnamefont {J.-Y.}\ \bibnamefont
  {Gyhm}}, \bibinfo {author} {\bibfnamefont {D.}~\bibnamefont {Rosa}},\ and\
  \bibinfo {author} {\bibfnamefont {D.}~\bibnamefont {\ifmmode~\check{S}\else
  \v{S}\fi{}afr\'anek}},\ }\bibfield  {title} {\bibinfo {title} {Minimal time
  required to charge a quantum system},\ }\href
  {https://doi.org/10.1103/PhysRevA.109.022607} {\bibfield  {journal} {\bibinfo
   {journal} {Phys. Rev. A}\ }\textbf {\bibinfo {volume} {109}},\ \bibinfo
  {pages} {022607} (\bibinfo {year} {2024})}\BibitemShut {NoStop}%
\bibitem [{\citenamefont {Fogarty}\ \emph {et~al.}(2020)\citenamefont
  {Fogarty}, \citenamefont {Deffner}, \citenamefont {Busch},\ and\
  \citenamefont {Campbell}}]{Fogarty2020PRL}%
  \BibitemOpen
  \bibfield  {author} {\bibinfo {author} {\bibfnamefont {T.}~\bibnamefont
  {Fogarty}}, \bibinfo {author} {\bibfnamefont {S.}~\bibnamefont {Deffner}},
  \bibinfo {author} {\bibfnamefont {T.}~\bibnamefont {Busch}},\ and\ \bibinfo
  {author} {\bibfnamefont {S.}~\bibnamefont {Campbell}},\ }\bibfield  {title}
  {\bibinfo {title} {{Orthogonality Catastrophe as a Consequence of the Quantum
  Speed Limit}},\ }\href {https://doi.org/10.1103/PhysRevLett.124.110601}
  {\bibfield  {journal} {\bibinfo  {journal} {Phys. Rev. Lett.}\ }\textbf
  {\bibinfo {volume} {124}},\ \bibinfo {pages} {110601} (\bibinfo {year}
  {2020})}\BibitemShut {NoStop}%
\bibitem [{\citenamefont {Girolami}\ and\ \citenamefont
  {Anz\`a}(2021)}]{Girolami2021PRL}%
  \BibitemOpen
  \bibfield  {author} {\bibinfo {author} {\bibfnamefont {D.}~\bibnamefont
  {Girolami}}\ and\ \bibinfo {author} {\bibfnamefont {F.}~\bibnamefont
  {Anz\`a}},\ }\bibfield  {title} {\bibinfo {title} {{Quantifying the
  Difference between Many-Body Quantum States}},\ }\href
  {https://doi.org/10.1103/PhysRevLett.126.170502} {\bibfield  {journal}
  {\bibinfo  {journal} {Phys. Rev. Lett.}\ }\textbf {\bibinfo {volume} {126}},\
  \bibinfo {pages} {170502} (\bibinfo {year} {2021})}\BibitemShut {NoStop}%
\bibitem [{\citenamefont {Mandelstam}\ and\ \citenamefont
  {Tamm}(1945)}]{MTBound_1945}%
  \BibitemOpen
  \bibfield  {author} {\bibinfo {author} {\bibfnamefont {L.}~\bibnamefont
  {Mandelstam}}\ and\ \bibinfo {author} {\bibfnamefont {I.}~\bibnamefont
  {Tamm}},\ }\bibfield  {title} {\bibinfo {title} {{The uncertainty relation
  between energy and time in nonrelativistic quantum mechanics}},\ }\href
  {https://link.springer.com/chapter/10.1007/978-3-642-74626-0_8} {\bibfield
  {journal} {\bibinfo  {journal} {J. Phys. (USSR)}\ }\textbf {\bibinfo {volume}
  {9}},\ \bibinfo {pages} {249} (\bibinfo {year} {1945})}\BibitemShut {NoStop}%
\bibitem [{\citenamefont {Margolus}\ and\ \citenamefont
  {Levitin}(1998)}]{Margolus1998}%
  \BibitemOpen
  \bibfield  {author} {\bibinfo {author} {\bibfnamefont {N.}~\bibnamefont
  {Margolus}}\ and\ \bibinfo {author} {\bibfnamefont {L.~B.}\ \bibnamefont
  {Levitin}},\ }\bibfield  {title} {\bibinfo {title} {The maximum speed of
  dynamical evolution},\ }\href
  {https://doi.org/https://doi.org/10.1016/S0167-2789(98)00054-2} {\bibfield
  {journal} {\bibinfo  {journal} {Physica D: Nonlinear Phenomena}\ }\textbf
  {\bibinfo {volume} {120}},\ \bibinfo {pages} {188} (\bibinfo {year}
  {1998})}\BibitemShut {NoStop}%
\bibitem [{\citenamefont {Levitin}\ and\ \citenamefont
  {Toffoli}(2009)}]{Levitin2009PRL}%
  \BibitemOpen
  \bibfield  {author} {\bibinfo {author} {\bibfnamefont {L.~B.}\ \bibnamefont
  {Levitin}}\ and\ \bibinfo {author} {\bibfnamefont {T.}~\bibnamefont
  {Toffoli}},\ }\bibfield  {title} {\bibinfo {title} {{Fundamental Limit on the
  Rate of Quantum Dynamics: The Unified Bound Is Tight}},\ }\href
  {https://doi.org/10.1103/PhysRevLett.103.160502} {\bibfield  {journal}
  {\bibinfo  {journal} {Phys. Rev. Lett.}\ }\textbf {\bibinfo {volume} {103}},\
  \bibinfo {pages} {160502} (\bibinfo {year} {2009})}\BibitemShut {NoStop}%
\bibitem [{\citenamefont {Anandan}\ and\ \citenamefont
  {Aharonov}(1990)}]{Anandan1990PRL}%
  \BibitemOpen
  \bibfield  {author} {\bibinfo {author} {\bibfnamefont {J.}~\bibnamefont
  {Anandan}}\ and\ \bibinfo {author} {\bibfnamefont {Y.}~\bibnamefont
  {Aharonov}},\ }\bibfield  {title} {\bibinfo {title} {Geometry of quantum
  evolution},\ }\href {https://doi.org/10.1103/PhysRevLett.65.1697} {\bibfield
  {journal} {\bibinfo  {journal} {Phys. Rev. Lett.}\ }\textbf {\bibinfo
  {volume} {65}},\ \bibinfo {pages} {1697} (\bibinfo {year}
  {1990})}\BibitemShut {NoStop}%
\bibitem [{\citenamefont {Pfeifer}(1993)}]{Pfeifer1993PRL}%
  \BibitemOpen
  \bibfield  {author} {\bibinfo {author} {\bibfnamefont {P.}~\bibnamefont
  {Pfeifer}},\ }\bibfield  {title} {\bibinfo {title} {How fast can a quantum
  state change with time?},\ }\href
  {https://doi.org/10.1103/PhysRevLett.70.3365} {\bibfield  {journal} {\bibinfo
   {journal} {Phys. Rev. Lett.}\ }\textbf {\bibinfo {volume} {70}},\ \bibinfo
  {pages} {3365} (\bibinfo {year} {1993})}\BibitemShut {NoStop}%
\bibitem [{\citenamefont {Uhlmann}(1992)}]{Uhlmann1992}%
  \BibitemOpen
  \bibfield  {author} {\bibinfo {author} {\bibfnamefont {A.}~\bibnamefont
  {Uhlmann}},\ }\bibfield  {title} {\bibinfo {title} {An energy dispersion
  estimate},\ }\href
  {https://doi.org/https://doi.org/10.1016/0375-9601(92)90555-Z} {\bibfield
  {journal} {\bibinfo  {journal} {Physics Letters A}\ }\textbf {\bibinfo
  {volume} {161}},\ \bibinfo {pages} {329} (\bibinfo {year}
  {1992})}\BibitemShut {NoStop}%
\bibitem [{\citenamefont {Marvian}\ \emph {et~al.}(2016)\citenamefont
  {Marvian}, \citenamefont {Spekkens},\ and\ \citenamefont
  {Zanardi}}]{Marvian2016PRA}%
  \BibitemOpen
  \bibfield  {author} {\bibinfo {author} {\bibfnamefont {I.}~\bibnamefont
  {Marvian}}, \bibinfo {author} {\bibfnamefont {R.~W.}\ \bibnamefont
  {Spekkens}},\ and\ \bibinfo {author} {\bibfnamefont {P.}~\bibnamefont
  {Zanardi}},\ }\bibfield  {title} {\bibinfo {title} {Quantum speed limits,
  coherence, and asymmetry},\ }\href
  {https://doi.org/10.1103/PhysRevA.93.052331} {\bibfield  {journal} {\bibinfo
  {journal} {Phys. Rev. A}\ }\textbf {\bibinfo {volume} {93}},\ \bibinfo
  {pages} {052331} (\bibinfo {year} {2016})}\BibitemShut {NoStop}%
\bibitem [{\citenamefont {Pires}\ \emph {et~al.}(2016)\citenamefont {Pires},
  \citenamefont {Cianciaruso}, \citenamefont {C\'eleri}, \citenamefont
  {Adesso},\ and\ \citenamefont {Soares-Pinto}}]{Pires2016PRX}%
  \BibitemOpen
  \bibfield  {author} {\bibinfo {author} {\bibfnamefont {D.~P.}\ \bibnamefont
  {Pires}}, \bibinfo {author} {\bibfnamefont {M.}~\bibnamefont {Cianciaruso}},
  \bibinfo {author} {\bibfnamefont {L.~C.}\ \bibnamefont {C\'eleri}}, \bibinfo
  {author} {\bibfnamefont {G.}~\bibnamefont {Adesso}},\ and\ \bibinfo {author}
  {\bibfnamefont {D.~O.}\ \bibnamefont {Soares-Pinto}},\ }\bibfield  {title}
  {\bibinfo {title} {{Generalized Geometric Quantum Speed Limits}},\ }\href
  {https://doi.org/10.1103/PhysRevX.6.021031} {\bibfield  {journal} {\bibinfo
  {journal} {Phys. Rev. X}\ }\textbf {\bibinfo {volume} {6}},\ \bibinfo {pages}
  {021031} (\bibinfo {year} {2016})}\BibitemShut {NoStop}%
\bibitem [{\citenamefont {Campaioli}\ \emph {et~al.}(2018)\citenamefont
  {Campaioli}, \citenamefont {Pollock}, \citenamefont {Binder},\ and\
  \citenamefont {Modi}}]{Campaioli2018PRL}%
  \BibitemOpen
  \bibfield  {author} {\bibinfo {author} {\bibfnamefont {F.}~\bibnamefont
  {Campaioli}}, \bibinfo {author} {\bibfnamefont {F.~A.}\ \bibnamefont
  {Pollock}}, \bibinfo {author} {\bibfnamefont {F.~C.}\ \bibnamefont
  {Binder}},\ and\ \bibinfo {author} {\bibfnamefont {K.}~\bibnamefont {Modi}},\
  }\bibfield  {title} {\bibinfo {title} {{Tightening Quantum Speed Limits for
  Almost All States}},\ }\href {https://doi.org/10.1103/PhysRevLett.120.060409}
  {\bibfield  {journal} {\bibinfo  {journal} {Phys. Rev. Lett.}\ }\textbf
  {\bibinfo {volume} {120}},\ \bibinfo {pages} {060409} (\bibinfo {year}
  {2018})}\BibitemShut {NoStop}%
\bibitem [{\citenamefont {Hörnedal}\ \emph {et~al.}(2022)\citenamefont
  {Hörnedal}, \citenamefont {Allan},\ and\ \citenamefont
  {Sönnerborn}}]{Niklas_2022_NJP}%
  \BibitemOpen
  \bibfield  {author} {\bibinfo {author} {\bibfnamefont {N.}~\bibnamefont
  {Hörnedal}}, \bibinfo {author} {\bibfnamefont {D.}~\bibnamefont {Allan}},\
  and\ \bibinfo {author} {\bibfnamefont {O.}~\bibnamefont {Sönnerborn}},\
  }\bibfield  {title} {\bibinfo {title} {{Extensions of the Mandelstam–Tamm
  quantum speed limit to systems in mixed states}},\ }\href
  {https://doi.org/10.1088/1367-2630/ac688a} {\bibfield  {journal} {\bibinfo
  {journal} {New Journal of Physics}\ }\textbf {\bibinfo {volume} {24}},\
  \bibinfo {pages} {055004} (\bibinfo {year} {2022})}\BibitemShut {NoStop}%
\bibitem [{\citenamefont {Rosal}\ \emph {et~al.}(2025)\citenamefont {Rosal},
  \citenamefont {Soares-Pinto},\ and\ \citenamefont
  {Pires}}]{Alberto_2024_PLA}%
  \BibitemOpen
  \bibfield  {author} {\bibinfo {author} {\bibfnamefont {A.~J.}\ \bibnamefont
  {Rosal}}, \bibinfo {author} {\bibfnamefont {D.~O.}\ \bibnamefont
  {Soares-Pinto}},\ and\ \bibinfo {author} {\bibfnamefont {D.~P.}\ \bibnamefont
  {Pires}},\ }\bibfield  {title} {\bibinfo {title} {{Quantum speed limits based
  on Schatten norms: Universality and tightness}},\ }\href
  {https://doi.org/https://doi.org/10.1016/j.physleta.2025.130250} {\bibfield
  {journal} {\bibinfo  {journal} {Physics Letters A}\ }\textbf {\bibinfo
  {volume} {534}},\ \bibinfo {pages} {130250} (\bibinfo {year}
  {2025})}\BibitemShut {NoStop}%
\bibitem [{\citenamefont {Deffner}\ and\ \citenamefont
  {Lutz}(2013)}]{Deffner2013PRL}%
  \BibitemOpen
  \bibfield  {author} {\bibinfo {author} {\bibfnamefont {S.}~\bibnamefont
  {Deffner}}\ and\ \bibinfo {author} {\bibfnamefont {E.}~\bibnamefont {Lutz}},\
  }\bibfield  {title} {\bibinfo {title} {{Quantum Speed Limit for Non-Markovian
  Dynamics}},\ }\href {https://doi.org/10.1103/PhysRevLett.111.010402}
  {\bibfield  {journal} {\bibinfo  {journal} {Phys. Rev. Lett.}\ }\textbf
  {\bibinfo {volume} {111}},\ \bibinfo {pages} {010402} (\bibinfo {year}
  {2013})}\BibitemShut {NoStop}%
\bibitem [{\citenamefont {del Campo}\ \emph {et~al.}(2013)\citenamefont {del
  Campo}, \citenamefont {Egusquiza}, \citenamefont {Plenio},\ and\
  \citenamefont {Huelga}}]{delCampo2013PRL}%
  \BibitemOpen
  \bibfield  {author} {\bibinfo {author} {\bibfnamefont {A.}~\bibnamefont {del
  Campo}}, \bibinfo {author} {\bibfnamefont {I.~L.}\ \bibnamefont {Egusquiza}},
  \bibinfo {author} {\bibfnamefont {M.~B.}\ \bibnamefont {Plenio}},\ and\
  \bibinfo {author} {\bibfnamefont {S.~F.}\ \bibnamefont {Huelga}},\ }\bibfield
   {title} {\bibinfo {title} {{Quantum Speed Limits in Open System Dynamics}},\
  }\href {https://doi.org/10.1103/PhysRevLett.110.050403} {\bibfield  {journal}
  {\bibinfo  {journal} {Phys. Rev. Lett.}\ }\textbf {\bibinfo {volume} {110}},\
  \bibinfo {pages} {050403} (\bibinfo {year} {2013})}\BibitemShut {NoStop}%
\bibitem [{\citenamefont {Taddei}\ \emph {et~al.}(2013)\citenamefont {Taddei},
  \citenamefont {Escher}, \citenamefont {Davidovich},\ and\ \citenamefont
  {de~Matos~Filho}}]{Taddei2013PRL}%
  \BibitemOpen
  \bibfield  {author} {\bibinfo {author} {\bibfnamefont {M.~M.}\ \bibnamefont
  {Taddei}}, \bibinfo {author} {\bibfnamefont {B.~M.}\ \bibnamefont {Escher}},
  \bibinfo {author} {\bibfnamefont {L.}~\bibnamefont {Davidovich}},\ and\
  \bibinfo {author} {\bibfnamefont {R.~L.}\ \bibnamefont {de~Matos~Filho}},\
  }\bibfield  {title} {\bibinfo {title} {{Quantum Speed Limit for Physical
  Processes}},\ }\href {https://doi.org/10.1103/PhysRevLett.110.050402}
  {\bibfield  {journal} {\bibinfo  {journal} {Phys. Rev. Lett.}\ }\textbf
  {\bibinfo {volume} {110}},\ \bibinfo {pages} {050402} (\bibinfo {year}
  {2013})}\BibitemShut {NoStop}%
\bibitem [{\citenamefont {Marvian}\ and\ \citenamefont
  {Lidar}(2015)}]{Marvian2015PRL}%
  \BibitemOpen
  \bibfield  {author} {\bibinfo {author} {\bibfnamefont {I.}~\bibnamefont
  {Marvian}}\ and\ \bibinfo {author} {\bibfnamefont {D.~A.}\ \bibnamefont
  {Lidar}},\ }\bibfield  {title} {\bibinfo {title} {{Quantum Speed Limits for
  Leakage and Decoherence}},\ }\href
  {https://doi.org/10.1103/PhysRevLett.115.210402} {\bibfield  {journal}
  {\bibinfo  {journal} {Phys. Rev. Lett.}\ }\textbf {\bibinfo {volume} {115}},\
  \bibinfo {pages} {210402} (\bibinfo {year} {2015})}\BibitemShut {NoStop}%
\bibitem [{\citenamefont {Giovannetti}\ \emph {et~al.}(2003)\citenamefont
  {Giovannetti}, \citenamefont {Lloyd},\ and\ \citenamefont
  {Maccone}}]{Vittorio2003PRA}%
  \BibitemOpen
  \bibfield  {author} {\bibinfo {author} {\bibfnamefont {V.}~\bibnamefont
  {Giovannetti}}, \bibinfo {author} {\bibfnamefont {S.}~\bibnamefont {Lloyd}},\
  and\ \bibinfo {author} {\bibfnamefont {L.}~\bibnamefont {Maccone}},\
  }\bibfield  {title} {\bibinfo {title} {Quantum limits to dynamical
  evolution},\ }\href {https://doi.org/10.1103/PhysRevA.67.052109} {\bibfield
  {journal} {\bibinfo  {journal} {Phys. Rev. A}\ }\textbf {\bibinfo {volume}
  {67}},\ \bibinfo {pages} {052109} (\bibinfo {year} {2003})}\BibitemShut
  {NoStop}%
\bibitem [{\citenamefont {Zander}\ \emph {et~al.}(2007)\citenamefont {Zander},
  \citenamefont {Plastino}, \citenamefont {Plastino},\ and\ \citenamefont
  {Casas}}]{Zander_2007}%
  \BibitemOpen
  \bibfield  {author} {\bibinfo {author} {\bibfnamefont {C.}~\bibnamefont
  {Zander}}, \bibinfo {author} {\bibfnamefont {A.~R.}\ \bibnamefont
  {Plastino}}, \bibinfo {author} {\bibfnamefont {A.}~\bibnamefont {Plastino}},\
  and\ \bibinfo {author} {\bibfnamefont {M.}~\bibnamefont {Casas}},\ }\bibfield
   {title} {\bibinfo {title} {Entanglement and the speed of evolution of
  multi-partite quantum systems},\ }\href
  {https://doi.org/10.1088/1751-8113/40/11/020} {\bibfield  {journal} {\bibinfo
   {journal} {Journal of Physics A: Mathematical and Theoretical}\ }\textbf
  {\bibinfo {volume} {40}},\ \bibinfo {pages} {2861} (\bibinfo {year}
  {2007})}\BibitemShut {NoStop}%
\bibitem [{\citenamefont {Garc\'{\i}a-Pintos}\ \emph
  {et~al.}(2022)\citenamefont {Garc\'{\i}a-Pintos}, \citenamefont {Nicholson},
  \citenamefont {Green}, \citenamefont {del Campo},\ and\ \citenamefont
  {Gorshkov}}]{delCampo2022PRX}%
  \BibitemOpen
  \bibfield  {author} {\bibinfo {author} {\bibfnamefont {L.~P.}\ \bibnamefont
  {Garc\'{\i}a-Pintos}}, \bibinfo {author} {\bibfnamefont {S.~B.}\ \bibnamefont
  {Nicholson}}, \bibinfo {author} {\bibfnamefont {J.~R.}\ \bibnamefont
  {Green}}, \bibinfo {author} {\bibfnamefont {A.}~\bibnamefont {del Campo}},\
  and\ \bibinfo {author} {\bibfnamefont {A.~V.}\ \bibnamefont {Gorshkov}},\
  }\bibfield  {title} {\bibinfo {title} {{Unifying Quantum and Classical Speed
  Limits on Observables}},\ }\href {https://doi.org/10.1103/PhysRevX.12.011038}
  {\bibfield  {journal} {\bibinfo  {journal} {Phys. Rev. X}\ }\textbf {\bibinfo
  {volume} {12}},\ \bibinfo {pages} {011038} (\bibinfo {year}
  {2022})}\BibitemShut {NoStop}%
\bibitem [{\citenamefont {Hamazaki}(2022)}]{Hamazaki2022PRXQuantum}%
  \BibitemOpen
  \bibfield  {author} {\bibinfo {author} {\bibfnamefont {R.}~\bibnamefont
  {Hamazaki}},\ }\bibfield  {title} {\bibinfo {title} {Speed limits for
  macroscopic transitions},\ }\href
  {https://doi.org/10.1103/PRXQuantum.3.020319} {\bibfield  {journal} {\bibinfo
   {journal} {PRX Quantum}\ }\textbf {\bibinfo {volume} {3}},\ \bibinfo {pages}
  {020319} (\bibinfo {year} {2022})}\BibitemShut {NoStop}%
\bibitem [{\citenamefont {Shanahan}\ \emph {et~al.}(2018)\citenamefont
  {Shanahan}, \citenamefont {Chenu}, \citenamefont {Margolus},\ and\
  \citenamefont {del Campo}}]{Shanahan2018PRL}%
  \BibitemOpen
  \bibfield  {author} {\bibinfo {author} {\bibfnamefont {B.}~\bibnamefont
  {Shanahan}}, \bibinfo {author} {\bibfnamefont {A.}~\bibnamefont {Chenu}},
  \bibinfo {author} {\bibfnamefont {N.}~\bibnamefont {Margolus}},\ and\
  \bibinfo {author} {\bibfnamefont {A.}~\bibnamefont {del Campo}},\ }\bibfield
  {title} {\bibinfo {title} {{Quantum Speed Limits across the
  Quantum-to-Classical Transition}},\ }\href
  {https://doi.org/10.1103/PhysRevLett.120.070401} {\bibfield  {journal}
  {\bibinfo  {journal} {Phys. Rev. Lett.}\ }\textbf {\bibinfo {volume} {120}},\
  \bibinfo {pages} {070401} (\bibinfo {year} {2018})}\BibitemShut {NoStop}%
\bibitem [{\citenamefont {Okuyama}\ and\ \citenamefont
  {Ohzeki}(2018)}]{Okuyama2018PRL}%
  \BibitemOpen
  \bibfield  {author} {\bibinfo {author} {\bibfnamefont {M.}~\bibnamefont
  {Okuyama}}\ and\ \bibinfo {author} {\bibfnamefont {M.}~\bibnamefont
  {Ohzeki}},\ }\bibfield  {title} {\bibinfo {title} {{Quantum Speed Limit is
  Not Quantum}},\ }\href {https://doi.org/10.1103/PhysRevLett.120.070402}
  {\bibfield  {journal} {\bibinfo  {journal} {Phys. Rev. Lett.}\ }\textbf
  {\bibinfo {volume} {120}},\ \bibinfo {pages} {070402} (\bibinfo {year}
  {2018})}\BibitemShut {NoStop}%
\bibitem [{\citenamefont {Shiraishi}\ \emph {et~al.}(2018)\citenamefont
  {Shiraishi}, \citenamefont {Funo},\ and\ \citenamefont
  {Saito}}]{Shiraishi2018PRL}%
  \BibitemOpen
  \bibfield  {author} {\bibinfo {author} {\bibfnamefont {N.}~\bibnamefont
  {Shiraishi}}, \bibinfo {author} {\bibfnamefont {K.}~\bibnamefont {Funo}},\
  and\ \bibinfo {author} {\bibfnamefont {K.}~\bibnamefont {Saito}},\ }\bibfield
   {title} {\bibinfo {title} {{Speed Limit for Classical Stochastic
  Processes}},\ }\href {https://doi.org/10.1103/PhysRevLett.121.070601}
  {\bibfield  {journal} {\bibinfo  {journal} {Phys. Rev. Lett.}\ }\textbf
  {\bibinfo {volume} {121}},\ \bibinfo {pages} {070601} (\bibinfo {year}
  {2018})}\BibitemShut {NoStop}%
\bibitem [{\citenamefont {Nicholson}\ \emph {et~al.}(2020)\citenamefont
  {Nicholson}, \citenamefont {Garc{\'\i}a-Pintos}, \citenamefont {del Campo},\
  and\ \citenamefont {Green}}]{Nicholson2020NP}%
  \BibitemOpen
  \bibfield  {author} {\bibinfo {author} {\bibfnamefont {S.~B.}\ \bibnamefont
  {Nicholson}}, \bibinfo {author} {\bibfnamefont {L.~P.}\ \bibnamefont
  {Garc{\'\i}a-Pintos}}, \bibinfo {author} {\bibfnamefont {A.}~\bibnamefont
  {del Campo}},\ and\ \bibinfo {author} {\bibfnamefont {J.~R.}\ \bibnamefont
  {Green}},\ }\bibfield  {title} {\bibinfo {title} {Time--information
  uncertainty relations in thermodynamics},\ }\href
  {https://doi.org/10.1038/s41567-020-0981-y} {\bibfield  {journal} {\bibinfo
  {journal} {Nature Physics}\ }\textbf {\bibinfo {volume} {16}},\ \bibinfo
  {pages} {1211} (\bibinfo {year} {2020})}\BibitemShut {NoStop}%
\bibitem [{\citenamefont {Cimmarusti}\ \emph {et~al.}(2015)\citenamefont
  {Cimmarusti}, \citenamefont {Yan}, \citenamefont {Patterson}, \citenamefont
  {Corcos}, \citenamefont {Orozco},\ and\ \citenamefont
  {Deffner}}]{Cimmarusti2015PRL}%
  \BibitemOpen
  \bibfield  {author} {\bibinfo {author} {\bibfnamefont {A.~D.}\ \bibnamefont
  {Cimmarusti}}, \bibinfo {author} {\bibfnamefont {Z.}~\bibnamefont {Yan}},
  \bibinfo {author} {\bibfnamefont {B.~D.}\ \bibnamefont {Patterson}}, \bibinfo
  {author} {\bibfnamefont {L.~P.}\ \bibnamefont {Corcos}}, \bibinfo {author}
  {\bibfnamefont {L.~A.}\ \bibnamefont {Orozco}},\ and\ \bibinfo {author}
  {\bibfnamefont {S.}~\bibnamefont {Deffner}},\ }\bibfield  {title} {\bibinfo
  {title} {{Environment-Assisted Speed-up of the Field Evolution in Cavity
  Quantum Electrodynamics}},\ }\href
  {https://doi.org/10.1103/PhysRevLett.114.233602} {\bibfield  {journal}
  {\bibinfo  {journal} {Phys. Rev. Lett.}\ }\textbf {\bibinfo {volume} {114}},\
  \bibinfo {pages} {233602} (\bibinfo {year} {2015})}\BibitemShut {NoStop}%
\bibitem [{\citenamefont {Lam}\ \emph {et~al.}(2021)\citenamefont {Lam},
  \citenamefont {Peter}, \citenamefont {Groh}, \citenamefont {Alt},
  \citenamefont {Robens}, \citenamefont {Meschede}, \citenamefont {Negretti},
  \citenamefont {Montangero}, \citenamefont {Calarco},\ and\ \citenamefont
  {Alberti}}]{Lam2021PRX}%
  \BibitemOpen
  \bibfield  {author} {\bibinfo {author} {\bibfnamefont {M.~R.}\ \bibnamefont
  {Lam}}, \bibinfo {author} {\bibfnamefont {N.}~\bibnamefont {Peter}}, \bibinfo
  {author} {\bibfnamefont {T.}~\bibnamefont {Groh}}, \bibinfo {author}
  {\bibfnamefont {W.}~\bibnamefont {Alt}}, \bibinfo {author} {\bibfnamefont
  {C.}~\bibnamefont {Robens}}, \bibinfo {author} {\bibfnamefont
  {D.}~\bibnamefont {Meschede}}, \bibinfo {author} {\bibfnamefont
  {A.}~\bibnamefont {Negretti}}, \bibinfo {author} {\bibfnamefont
  {S.}~\bibnamefont {Montangero}}, \bibinfo {author} {\bibfnamefont
  {T.}~\bibnamefont {Calarco}},\ and\ \bibinfo {author} {\bibfnamefont
  {A.}~\bibnamefont {Alberti}},\ }\bibfield  {title} {\bibinfo {title}
  {{Demonstration of Quantum Brachistochrones between Distant States of an
  Atom}},\ }\href {https://doi.org/10.1103/PhysRevX.11.011035} {\bibfield
  {journal} {\bibinfo  {journal} {Phys. Rev. X}\ }\textbf {\bibinfo {volume}
  {11}},\ \bibinfo {pages} {011035} (\bibinfo {year} {2021})}\BibitemShut
  {NoStop}%
\bibitem [{\citenamefont {Ness}\ \emph {et~al.}(2021)\citenamefont {Ness},
  \citenamefont {Lam}, \citenamefont {Alt}, \citenamefont {Meschede},
  \citenamefont {Sagi},\ and\ \citenamefont {Alberti}}]{Gal2021SciAdv}%
  \BibitemOpen
  \bibfield  {author} {\bibinfo {author} {\bibfnamefont {G.}~\bibnamefont
  {Ness}}, \bibinfo {author} {\bibfnamefont {M.~R.}\ \bibnamefont {Lam}},
  \bibinfo {author} {\bibfnamefont {W.}~\bibnamefont {Alt}}, \bibinfo {author}
  {\bibfnamefont {D.}~\bibnamefont {Meschede}}, \bibinfo {author}
  {\bibfnamefont {Y.}~\bibnamefont {Sagi}},\ and\ \bibinfo {author}
  {\bibfnamefont {A.}~\bibnamefont {Alberti}},\ }\bibfield  {title} {\bibinfo
  {title} {Observing crossover between quantum speed limits},\ }\href
  {https://doi.org/10.1126/sciadv.abj9119} {\bibfield  {journal} {\bibinfo
  {journal} {Science Advances}\ }\textbf {\bibinfo {volume} {7}},\ \bibinfo
  {pages} {eabj9119} (\bibinfo {year} {2021})}\BibitemShut {NoStop}%
\bibitem [{\citenamefont {Streltsov}\ \emph {et~al.}(2017)\citenamefont
  {Streltsov}, \citenamefont {Adesso},\ and\ \citenamefont
  {Plenio}}]{Plenio2017RMP}%
  \BibitemOpen
  \bibfield  {author} {\bibinfo {author} {\bibfnamefont {A.}~\bibnamefont
  {Streltsov}}, \bibinfo {author} {\bibfnamefont {G.}~\bibnamefont {Adesso}},\
  and\ \bibinfo {author} {\bibfnamefont {M.~B.}\ \bibnamefont {Plenio}},\
  }\bibfield  {title} {\bibinfo {title} {{Colloquium: Quantum coherence as a
  resource}},\ }\href {https://doi.org/10.1103/RevModPhys.89.041003} {\bibfield
   {journal} {\bibinfo  {journal} {Rev. Mod. Phys.}\ }\textbf {\bibinfo
  {volume} {89}},\ \bibinfo {pages} {041003} (\bibinfo {year}
  {2017})}\BibitemShut {NoStop}%
\bibitem [{\citenamefont {Chitambar}\ and\ \citenamefont
  {Hsieh}(2016)}]{Chitambar_2016_PRL}%
  \BibitemOpen
  \bibfield  {author} {\bibinfo {author} {\bibfnamefont {E.}~\bibnamefont
  {Chitambar}}\ and\ \bibinfo {author} {\bibfnamefont {M.-H.}\ \bibnamefont
  {Hsieh}},\ }\bibfield  {title} {\bibinfo {title} {{Relating the Resource
  Theories of Entanglement and Quantum Coherence}},\ }\href
  {https://doi.org/10.1103/PhysRevLett.117.020402} {\bibfield  {journal}
  {\bibinfo  {journal} {Phys. Rev. Lett.}\ }\textbf {\bibinfo {volume} {117}},\
  \bibinfo {pages} {020402} (\bibinfo {year} {2016})}\BibitemShut {NoStop}%
\bibitem [{\citenamefont {Tan}\ \emph {et~al.}(2016)\citenamefont {Tan},
  \citenamefont {Kwon}, \citenamefont {Park},\ and\ \citenamefont
  {Jeong}}]{Tan_2016_PRA}%
  \BibitemOpen
  \bibfield  {author} {\bibinfo {author} {\bibfnamefont {K.~C.}\ \bibnamefont
  {Tan}}, \bibinfo {author} {\bibfnamefont {H.}~\bibnamefont {Kwon}}, \bibinfo
  {author} {\bibfnamefont {C.-Y.}\ \bibnamefont {Park}},\ and\ \bibinfo
  {author} {\bibfnamefont {H.}~\bibnamefont {Jeong}},\ }\bibfield  {title}
  {\bibinfo {title} {Unified view of quantum correlations and quantum
  coherence},\ }\href {https://doi.org/10.1103/PhysRevA.94.022329} {\bibfield
  {journal} {\bibinfo  {journal} {Phys. Rev. A}\ }\textbf {\bibinfo {volume}
  {94}},\ \bibinfo {pages} {022329} (\bibinfo {year} {2016})}\BibitemShut
  {NoStop}%
\bibitem [{\citenamefont {Pires}\ \emph {et~al.}(2015)\citenamefont {Pires},
  \citenamefont {C\'eleri},\ and\ \citenamefont {Soares-Pinto}}]{Pires2015PRA}%
  \BibitemOpen
  \bibfield  {author} {\bibinfo {author} {\bibfnamefont {D.~P.}\ \bibnamefont
  {Pires}}, \bibinfo {author} {\bibfnamefont {L.~C.}\ \bibnamefont
  {C\'eleri}},\ and\ \bibinfo {author} {\bibfnamefont {D.~O.}\ \bibnamefont
  {Soares-Pinto}},\ }\bibfield  {title} {\bibinfo {title} {Geometric lower
  bound for a quantum coherence measure},\ }\href
  {https://doi.org/10.1103/PhysRevA.91.042330} {\bibfield  {journal} {\bibinfo
  {journal} {Phys. Rev. A}\ }\textbf {\bibinfo {volume} {91}},\ \bibinfo
  {pages} {042330} (\bibinfo {year} {2015})}\BibitemShut {NoStop}%
\bibitem [{\citenamefont {Jing}\ \emph {et~al.}(2016)\citenamefont {Jing},
  \citenamefont {Wu},\ and\ \citenamefont {del Campo}}]{delCampo_2016_SRep}%
  \BibitemOpen
  \bibfield  {author} {\bibinfo {author} {\bibfnamefont {J.}~\bibnamefont
  {Jing}}, \bibinfo {author} {\bibfnamefont {L.-A.}\ \bibnamefont {Wu}},\ and\
  \bibinfo {author} {\bibfnamefont {A.}~\bibnamefont {del Campo}},\ }\bibfield
  {title} {\bibinfo {title} {{Fundamental Speed Limits to the Generation of
  Quantumness}},\ }\href {https://doi.org/10.1038/srep38149} {\bibfield
  {journal} {\bibinfo  {journal} {Scientific Reports}\ }\textbf {\bibinfo
  {volume} {6}},\ \bibinfo {pages} {38149} (\bibinfo {year}
  {2016})}\BibitemShut {NoStop}%
\bibitem [{\citenamefont {Funo}\ \emph {et~al.}(2017)\citenamefont {Funo},
  \citenamefont {Zhang}, \citenamefont {Chatou}, \citenamefont {Kim},
  \citenamefont {Ueda},\ and\ \citenamefont {del Campo}}]{delCampo_2017_PRL}%
  \BibitemOpen
  \bibfield  {author} {\bibinfo {author} {\bibfnamefont {K.}~\bibnamefont
  {Funo}}, \bibinfo {author} {\bibfnamefont {J.-N.}\ \bibnamefont {Zhang}},
  \bibinfo {author} {\bibfnamefont {C.}~\bibnamefont {Chatou}}, \bibinfo
  {author} {\bibfnamefont {K.}~\bibnamefont {Kim}}, \bibinfo {author}
  {\bibfnamefont {M.}~\bibnamefont {Ueda}},\ and\ \bibinfo {author}
  {\bibfnamefont {A.}~\bibnamefont {del Campo}},\ }\bibfield  {title} {\bibinfo
  {title} {{Universal Work Fluctuations During Shortcuts to Adiabaticity by
  Counterdiabatic Driving}},\ }\href
  {https://doi.org/10.1103/PhysRevLett.118.100602} {\bibfield  {journal}
  {\bibinfo  {journal} {Phys. Rev. Lett.}\ }\textbf {\bibinfo {volume} {118}},\
  \bibinfo {pages} {100602} (\bibinfo {year} {2017})}\BibitemShut {NoStop}%
\bibitem [{\citenamefont {Rossatto}\ \emph {et~al.}(2020)\citenamefont
  {Rossatto}, \citenamefont {Pires}, \citenamefont {de~Paula},\ and\
  \citenamefont {de~S\'a~Neto}}]{Rossatto2020PRA}%
  \BibitemOpen
  \bibfield  {author} {\bibinfo {author} {\bibfnamefont {D.~Z.}\ \bibnamefont
  {Rossatto}}, \bibinfo {author} {\bibfnamefont {D.~P.}\ \bibnamefont {Pires}},
  \bibinfo {author} {\bibfnamefont {F.~M.}\ \bibnamefont {de~Paula}},\ and\
  \bibinfo {author} {\bibfnamefont {O.~P.}\ \bibnamefont {de~S\'a~Neto}},\
  }\bibfield  {title} {\bibinfo {title} {{Quantum coherence and speed limit in
  the mean-field Dicke model of superradiance}},\ }\href
  {https://doi.org/10.1103/PhysRevA.102.053716} {\bibfield  {journal} {\bibinfo
   {journal} {Phys. Rev. A}\ }\textbf {\bibinfo {volume} {102}},\ \bibinfo
  {pages} {053716} (\bibinfo {year} {2020})}\BibitemShut {NoStop}%
\bibitem [{\citenamefont {Paulson}\ and\ \citenamefont
  {Banerjee}(2022)}]{Paulson_2022}%
  \BibitemOpen
  \bibfield  {author} {\bibinfo {author} {\bibfnamefont {K.~G.}\ \bibnamefont
  {Paulson}}\ and\ \bibinfo {author} {\bibfnamefont {S.}~\bibnamefont
  {Banerjee}},\ }\bibfield  {title} {\bibinfo {title} {{Quantum speed limit
  time: role of coherence}},\ }\href {https://doi.org/10.1088/1751-8121/acaadb}
  {\bibfield  {journal} {\bibinfo  {journal} {Journal of Physics A:
  Mathematical and Theoretical}\ }\textbf {\bibinfo {volume} {55}},\ \bibinfo
  {pages} {505302} (\bibinfo {year} {2022})}\BibitemShut {NoStop}%
\bibitem [{\citenamefont {Baumgratz}\ \emph {et~al.}(2014)\citenamefont
  {Baumgratz}, \citenamefont {Cramer},\ and\ \citenamefont
  {Plenio}}]{Plenio2014PRL}%
  \BibitemOpen
  \bibfield  {author} {\bibinfo {author} {\bibfnamefont {T.}~\bibnamefont
  {Baumgratz}}, \bibinfo {author} {\bibfnamefont {M.}~\bibnamefont {Cramer}},\
  and\ \bibinfo {author} {\bibfnamefont {M.~B.}\ \bibnamefont {Plenio}},\
  }\bibfield  {title} {\bibinfo {title} {{Quantifying Coherence}},\ }\href
  {https://doi.org/10.1103/PhysRevLett.113.140401} {\bibfield  {journal}
  {\bibinfo  {journal} {Phys. Rev. Lett.}\ }\textbf {\bibinfo {volume} {113}},\
  \bibinfo {pages} {140401} (\bibinfo {year} {2014})}\BibitemShut {NoStop}%
\bibitem [{\citenamefont {de~Vicente}\ and\ \citenamefont
  {Streltsov}(2016)}]{deVicente_2017}%
  \BibitemOpen
  \bibfield  {author} {\bibinfo {author} {\bibfnamefont {J.~I.}\ \bibnamefont
  {de~Vicente}}\ and\ \bibinfo {author} {\bibfnamefont {A.}~\bibnamefont
  {Streltsov}},\ }\bibfield  {title} {\bibinfo {title} {Genuine quantum
  coherence},\ }\href {https://doi.org/10.1088/1751-8121/50/4/045301}
  {\bibfield  {journal} {\bibinfo  {journal} {Journal of Physics A:
  Mathematical and Theoretical}\ }\textbf {\bibinfo {volume} {50}},\ \bibinfo
  {pages} {045301} (\bibinfo {year} {2016})}\BibitemShut {NoStop}%
\bibitem [{\citenamefont {Jin}\ and\ \citenamefont {Fei}(2018)}]{Jin2018PRA}%
  \BibitemOpen
  \bibfield  {author} {\bibinfo {author} {\bibfnamefont {Z.-X.}\ \bibnamefont
  {Jin}}\ and\ \bibinfo {author} {\bibfnamefont {S.-M.}\ \bibnamefont {Fei}},\
  }\bibfield  {title} {\bibinfo {title} {{Quantifying quantum coherence and
  nonclassical correlation based on Hellinger distance}},\ }\href
  {https://doi.org/10.1103/PhysRevA.97.062342} {\bibfield  {journal} {\bibinfo
  {journal} {Phys. Rev. A}\ }\textbf {\bibinfo {volume} {97}},\ \bibinfo
  {pages} {062342} (\bibinfo {year} {2018})}\BibitemShut {NoStop}%
\bibitem [{\citenamefont {Ivakhnenko}\ \emph {et~al.}(2023)\citenamefont
  {Ivakhnenko}, \citenamefont {Shevchenko},\ and\ \citenamefont
  {Nori}}]{Ivakhnenko_2023}%
  \BibitemOpen
  \bibfield  {author} {\bibinfo {author} {\bibfnamefont {O.~V.}\ \bibnamefont
  {Ivakhnenko}}, \bibinfo {author} {\bibfnamefont {S.~N.}\ \bibnamefont
  {Shevchenko}},\ and\ \bibinfo {author} {\bibfnamefont {F.}~\bibnamefont
  {Nori}},\ }\bibfield  {title} {\bibinfo {title} {{Nonadiabatic
  Landau–Zener–Stückelberg–Majorana transitions, dynamics, and
  interference}},\ }\href
  {https://doi.org/https://doi.org/10.1016/j.physrep.2022.10.002} {\bibfield
  {journal} {\bibinfo  {journal} {Physics Reports}\ }\textbf {\bibinfo {volume}
  {995}},\ \bibinfo {pages} {1} (\bibinfo {year} {2023})}\BibitemShut {NoStop}%
\bibitem [{\citenamefont {Berry}(2009)}]{Berry_2009}%
  \BibitemOpen
  \bibfield  {author} {\bibinfo {author} {\bibfnamefont {M.~V.}\ \bibnamefont
  {Berry}},\ }\bibfield  {title} {\bibinfo {title} {Transitionless quantum
  driving},\ }\href {https://doi.org/10.1088/1751-8113/42/36/365303} {\bibfield
   {journal} {\bibinfo  {journal} {Journal of Physics A: Mathematical and
  Theoretical}\ }\textbf {\bibinfo {volume} {42}},\ \bibinfo {pages} {365303}
  (\bibinfo {year} {2009})}\BibitemShut {NoStop}%
\bibitem [{\citenamefont {Gu\'ery-Odelin}\ \emph {et~al.}(2019)\citenamefont
  {Gu\'ery-Odelin}, \citenamefont {Ruschhaupt}, \citenamefont {Kiely},
  \citenamefont {Torrontegui}, \citenamefont {Mart\'{\i}nez-Garaot},\ and\
  \citenamefont {Muga}}]{Muga_2019_RMP}%
  \BibitemOpen
  \bibfield  {author} {\bibinfo {author} {\bibfnamefont {D.}~\bibnamefont
  {Gu\'ery-Odelin}}, \bibinfo {author} {\bibfnamefont {A.}~\bibnamefont
  {Ruschhaupt}}, \bibinfo {author} {\bibfnamefont {A.}~\bibnamefont {Kiely}},
  \bibinfo {author} {\bibfnamefont {E.}~\bibnamefont {Torrontegui}}, \bibinfo
  {author} {\bibfnamefont {S.}~\bibnamefont {Mart\'{\i}nez-Garaot}},\ and\
  \bibinfo {author} {\bibfnamefont {J.~G.}\ \bibnamefont {Muga}},\ }\bibfield
  {title} {\bibinfo {title} {{Shortcuts to adiabaticity: Concepts, methods, and
  applications}},\ }\href {https://doi.org/10.1103/RevModPhys.91.045001}
  {\bibfield  {journal} {\bibinfo  {journal} {Rev. Mod. Phys.}\ }\textbf
  {\bibinfo {volume} {91}},\ \bibinfo {pages} {045001} (\bibinfo {year}
  {2019})}\BibitemShut {NoStop}%
\bibitem [{\citenamefont {Audenaert}(2014)}]{Audenaert2014}%
  \BibitemOpen
  \bibfield  {author} {\bibinfo {author} {\bibfnamefont {K.~M.~R.}\
  \bibnamefont {Audenaert}},\ }\bibfield  {title} {\bibinfo {title}
  {{Comparisons between Quantum State Distinguishability Measures}},\ }\href
  {https://dl.acm.org/doi/10.5555/2600498.2600500} {\bibfield  {journal}
  {\bibinfo  {journal} {Quantum Info. Comput.}\ }\textbf {\bibinfo {volume}
  {14}},\ \bibinfo {pages} {31–38} (\bibinfo {year} {2014})}\BibitemShut
  {NoStop}%
\bibitem [{\citenamefont {Luo}\ and\ \citenamefont {Zhang}(2004)}]{Luo2004PRA}%
  \BibitemOpen
  \bibfield  {author} {\bibinfo {author} {\bibfnamefont {S.}~\bibnamefont
  {Luo}}\ and\ \bibinfo {author} {\bibfnamefont {Q.}~\bibnamefont {Zhang}},\
  }\bibfield  {title} {\bibinfo {title} {Informational distance on
  quantum-state space},\ }\href {https://doi.org/10.1103/PhysRevA.69.032106}
  {\bibfield  {journal} {\bibinfo  {journal} {Phys. Rev. A}\ }\textbf {\bibinfo
  {volume} {69}},\ \bibinfo {pages} {032106} (\bibinfo {year}
  {2004})}\BibitemShut {NoStop}%
\bibitem [{\citenamefont {Nielsen}\ and\ \citenamefont
  {Chuang}(2010)}]{nielsen_chuang_2010}%
  \BibitemOpen
  \bibfield  {author} {\bibinfo {author} {\bibfnamefont {M.~A.}\ \bibnamefont
  {Nielsen}}\ and\ \bibinfo {author} {\bibfnamefont {I.~L.}\ \bibnamefont
  {Chuang}},\ }\href {https://doi.org/10.1017/CBO9780511976667} {\emph
  {\bibinfo {title} {Quantum Computation and Quantum Information: 10th
  Anniversary Edition}}}\ (\bibinfo  {publisher} {Cambridge University Press,
  Cambridge, England},\ \bibinfo {year} {2010})\BibitemShut {NoStop}%
\bibitem [{\citenamefont {Wigner}\ and\ \citenamefont
  {Yanase}(1963)}]{Wigner_Yanase_1963}%
  \BibitemOpen
  \bibfield  {author} {\bibinfo {author} {\bibfnamefont {E.~P.}\ \bibnamefont
  {Wigner}}\ and\ \bibinfo {author} {\bibfnamefont {M.~M.}\ \bibnamefont
  {Yanase}},\ }\bibfield  {title} {\bibinfo {title} {{Information Contents of
  Distributions}},\ }\href {https://doi.org/10.1073/pnas.49.6.910} {\bibfield
  {journal} {\bibinfo  {journal} {Proc. Nat. Acad. Sci.}\ }\textbf {\bibinfo
  {volume} {49}},\ \bibinfo {pages} {910} (\bibinfo {year} {1963})}\BibitemShut
  {NoStop}%
\bibitem [{\citenamefont {del Campo}\ \emph {et~al.}(2012)\citenamefont {del
  Campo}, \citenamefont {Rams},\ and\ \citenamefont
  {Zurek}}]{delCampo_2012_PRL}%
  \BibitemOpen
  \bibfield  {author} {\bibinfo {author} {\bibfnamefont {A.}~\bibnamefont {del
  Campo}}, \bibinfo {author} {\bibfnamefont {M.~M.}\ \bibnamefont {Rams}},\
  and\ \bibinfo {author} {\bibfnamefont {W.~H.}\ \bibnamefont {Zurek}},\
  }\bibfield  {title} {\bibinfo {title} {{Assisted Finite-Rate Adiabatic
  Passage Across a Quantum Critical Point: Exact Solution for the Quantum Ising
  Model}},\ }\href {https://doi.org/10.1103/PhysRevLett.109.115703} {\bibfield
  {journal} {\bibinfo  {journal} {Phys. Rev. Lett.}\ }\textbf {\bibinfo
  {volume} {109}},\ \bibinfo {pages} {115703} (\bibinfo {year}
  {2012})}\BibitemShut {NoStop}%
\bibitem [{\citenamefont {Johnson}\ \emph {et~al.}(2011)\citenamefont
  {Johnson}, \citenamefont {Amin}, \citenamefont {Gildert}, \citenamefont
  {Lanting}, \citenamefont {Hamze}, \citenamefont {Dickson}, \citenamefont
  {Harris}, \citenamefont {Berkley}, \citenamefont {Johansson}, \citenamefont
  {Bunyk}, \citenamefont {Chapple}, \citenamefont {Enderud}, \citenamefont
  {Hilton}, \citenamefont {Karimi}, \citenamefont {Ladizinsky}, \citenamefont
  {Ladizinsky}, \citenamefont {Oh}, \citenamefont {Perminov}, \citenamefont
  {Rich}, \citenamefont {Thom}, \citenamefont {Tolkacheva}, \citenamefont
  {Truncik}, \citenamefont {Uchaikin}, \citenamefont {Wang}, \citenamefont
  {Wilson},\ and\ \citenamefont {Rose}}]{Johnson_2011_Nature}%
  \BibitemOpen
  \bibfield  {author} {\bibinfo {author} {\bibfnamefont {M.~W.}\ \bibnamefont
  {Johnson}}, \bibinfo {author} {\bibfnamefont {M.~H.~S.}\ \bibnamefont
  {Amin}}, \bibinfo {author} {\bibfnamefont {S.}~\bibnamefont {Gildert}},
  \bibinfo {author} {\bibfnamefont {T.}~\bibnamefont {Lanting}}, \bibinfo
  {author} {\bibfnamefont {F.}~\bibnamefont {Hamze}}, \bibinfo {author}
  {\bibfnamefont {N.}~\bibnamefont {Dickson}}, \bibinfo {author} {\bibfnamefont
  {R.}~\bibnamefont {Harris}}, \bibinfo {author} {\bibfnamefont {A.~J.}\
  \bibnamefont {Berkley}}, \bibinfo {author} {\bibfnamefont {J.}~\bibnamefont
  {Johansson}}, \bibinfo {author} {\bibfnamefont {P.}~\bibnamefont {Bunyk}},
  \bibinfo {author} {\bibfnamefont {E.~M.}\ \bibnamefont {Chapple}}, \bibinfo
  {author} {\bibfnamefont {C.}~\bibnamefont {Enderud}}, \bibinfo {author}
  {\bibfnamefont {J.~P.}\ \bibnamefont {Hilton}}, \bibinfo {author}
  {\bibfnamefont {K.}~\bibnamefont {Karimi}}, \bibinfo {author} {\bibfnamefont
  {E.}~\bibnamefont {Ladizinsky}}, \bibinfo {author} {\bibfnamefont
  {N.}~\bibnamefont {Ladizinsky}}, \bibinfo {author} {\bibfnamefont
  {T.}~\bibnamefont {Oh}}, \bibinfo {author} {\bibfnamefont {I.}~\bibnamefont
  {Perminov}}, \bibinfo {author} {\bibfnamefont {C.}~\bibnamefont {Rich}},
  \bibinfo {author} {\bibfnamefont {M.~C.}\ \bibnamefont {Thom}}, \bibinfo
  {author} {\bibfnamefont {E.}~\bibnamefont {Tolkacheva}}, \bibinfo {author}
  {\bibfnamefont {C.~J.~S.}\ \bibnamefont {Truncik}}, \bibinfo {author}
  {\bibfnamefont {S.}~\bibnamefont {Uchaikin}}, \bibinfo {author}
  {\bibfnamefont {J.}~\bibnamefont {Wang}}, \bibinfo {author} {\bibfnamefont
  {B.}~\bibnamefont {Wilson}},\ and\ \bibinfo {author} {\bibfnamefont
  {G.}~\bibnamefont {Rose}},\ }\bibfield  {title} {\bibinfo {title} {{Quantum
  annealing with manufactured spins}},\ }\href
  {https://doi.org/10.1038/nature10012} {\bibfield  {journal} {\bibinfo
  {journal} {Nature}\ }\textbf {\bibinfo {volume} {473}},\ \bibinfo {pages}
  {194} (\bibinfo {year} {2011})}\BibitemShut {NoStop}%
\bibitem [{\citenamefont {Scholl}\ \emph {et~al.}(2021)\citenamefont {Scholl},
  \citenamefont {Schuler}, \citenamefont {Williams}, \citenamefont
  {Eberharter}, \citenamefont {Barredo}, \citenamefont {Schymik}, \citenamefont
  {Lienhard}, \citenamefont {Henry}, \citenamefont {Lang}, \citenamefont
  {Lahaye}, \citenamefont {L{\"a}uchli},\ and\ \citenamefont
  {Browaeys}}]{Scholl_2021_Nature}%
  \BibitemOpen
  \bibfield  {author} {\bibinfo {author} {\bibfnamefont {P.}~\bibnamefont
  {Scholl}}, \bibinfo {author} {\bibfnamefont {M.}~\bibnamefont {Schuler}},
  \bibinfo {author} {\bibfnamefont {H.~J.}\ \bibnamefont {Williams}}, \bibinfo
  {author} {\bibfnamefont {A.~A.}\ \bibnamefont {Eberharter}}, \bibinfo
  {author} {\bibfnamefont {D.}~\bibnamefont {Barredo}}, \bibinfo {author}
  {\bibfnamefont {K.-N.}\ \bibnamefont {Schymik}}, \bibinfo {author}
  {\bibfnamefont {V.}~\bibnamefont {Lienhard}}, \bibinfo {author}
  {\bibfnamefont {L.-P.}\ \bibnamefont {Henry}}, \bibinfo {author}
  {\bibfnamefont {T.~C.}\ \bibnamefont {Lang}}, \bibinfo {author}
  {\bibfnamefont {T.}~\bibnamefont {Lahaye}}, \bibinfo {author} {\bibfnamefont
  {A.~M.}\ \bibnamefont {L{\"a}uchli}},\ and\ \bibinfo {author} {\bibfnamefont
  {A.}~\bibnamefont {Browaeys}},\ }\bibfield  {title} {\bibinfo {title}
  {{Quantum simulation of 2D antiferromagnets with hundreds of Rydberg
  atoms}},\ }\href {https://doi.org/10.1038/s41586-021-03585-1} {\bibfield
  {journal} {\bibinfo  {journal} {Nature}\ }\textbf {\bibinfo {volume} {595}},\
  \bibinfo {pages} {233} (\bibinfo {year} {2021})}\BibitemShut {NoStop}%
\bibitem [{\citenamefont {Li}\ \emph {et~al.}(2023)\citenamefont {Li},
  \citenamefont {Wu}, \citenamefont {Mei}, \citenamefont {Yao}, \citenamefont
  {Lian}, \citenamefont {Cai}, \citenamefont {Wang}, \citenamefont {Qi},
  \citenamefont {Yao}, \citenamefont {He}, \citenamefont {Zhou},\ and\
  \citenamefont {Duan}}]{Duan_2023_PRXQuantum}%
  \BibitemOpen
  \bibfield  {author} {\bibinfo {author} {\bibfnamefont {B.-W.}\ \bibnamefont
  {Li}}, \bibinfo {author} {\bibfnamefont {Y.-K.}\ \bibnamefont {Wu}}, \bibinfo
  {author} {\bibfnamefont {Q.-X.}\ \bibnamefont {Mei}}, \bibinfo {author}
  {\bibfnamefont {R.}~\bibnamefont {Yao}}, \bibinfo {author} {\bibfnamefont
  {W.-Q.}\ \bibnamefont {Lian}}, \bibinfo {author} {\bibfnamefont {M.-L.}\
  \bibnamefont {Cai}}, \bibinfo {author} {\bibfnamefont {Y.}~\bibnamefont
  {Wang}}, \bibinfo {author} {\bibfnamefont {B.-X.}\ \bibnamefont {Qi}},
  \bibinfo {author} {\bibfnamefont {L.}~\bibnamefont {Yao}}, \bibinfo {author}
  {\bibfnamefont {L.}~\bibnamefont {He}}, \bibinfo {author} {\bibfnamefont
  {Z.-C.}\ \bibnamefont {Zhou}},\ and\ \bibinfo {author} {\bibfnamefont
  {L.-M.}\ \bibnamefont {Duan}},\ }\bibfield  {title} {\bibinfo {title}
  {{Probing Critical Behavior of Long-Range Transverse-Field Ising Model
  through Quantum Kibble-Zurek Mechanism}},\ }\href
  {https://doi.org/10.1103/PRXQuantum.4.010302} {\bibfield  {journal} {\bibinfo
   {journal} {PRX Quantum}\ }\textbf {\bibinfo {volume} {4}},\ \bibinfo {pages}
  {010302} (\bibinfo {year} {2023})}\BibitemShut {NoStop}%
\bibitem [{\citenamefont {Bhatia}(1997)}]{Bhatia2013matrix}%
  \BibitemOpen
  \bibfield  {author} {\bibinfo {author} {\bibfnamefont {R.}~\bibnamefont
  {Bhatia}},\ }\href {https://doi.org/10.1007/978-1-4612-0653-8} {\emph
  {\bibinfo {title} {{Matrix Analysis}}}}\ (\bibinfo  {publisher} {Springer New
  York, NY},\ \bibinfo {year} {1997})\BibitemShut {NoStop}%
\end{thebibliography}%

\end{document}